\documentclass[journal,10pt,twocolumn]{IEEEtran}

\usepackage[utf8]{inputenc}
\usepackage[T1]{fontenc}

\usepackage[english]{babel}
\usepackage{amsbsy,amsmath,amsfonts,amssymb,amsthm}
\usepackage{mathtools}
\usepackage{textcomp} 

\usepackage[noadjust]{cite}
\usepackage{bm,cases,url,verbatim} 
\usepackage{booktabs}

\usepackage{algorithm}
\usepackage[noend]{algpseudocode}

\usepackage{centernot} 

\newcommand{\subparagraph}{}

\newcommand{\argmin}{\operatornamewithlimits{argmin}}

\begingroup
\catcode`\#=11

\endgroup

\usepackage{graphicx,xcolor,float,dblfloatfix}
\usepackage{psfrag}

\usepackage{caption}
\usepackage{subcaption}

\newtheorem{thm}{Theorem}

\newcommand{\subalign}[1]{%
  \vcenter{%
    \Let@ \restore@math@cr \default@tag
    \baselineskip\fontdimen10 \scriptfont\tw@
    \advance\baselineskip\fontdimen12 \scriptfont\tw@
    \lineskip\thr@@\fontdimen8 \scriptfont\thr@@
    \lineskiplimit\lineskip
    \ialign{\hfil$\m@th\textstyle##$&$\m@th\textstyle{}##$\crcr
      #1\crcr
    }%
  }
}

\usepackage{epstopdf}

\graphicspath{%
{./}
}

\usepackage{pgfplots}
\pgfplotsset{compat=newest}
\usetikzlibrary{plotmarks}
\usetikzlibrary{arrows, decorations.markings,shapes,positioning,chains}
\usepgfplotslibrary{patchplots}
\usepackage{grffile}

\pgfplotsset{plot coordinates/math parser=false}
\newlength\fheight
\newlength\fwidth

\usepackage[flushleft]{threeparttable}

\newcommand{\ttx}{{\rm tx}}
\newcommand{\pr}{{\rm P}} 
\newcommand{\prr}{{\rm Pr}} 
\newcommand{\out}{{\rm out}} 
\newcommand{\fl}{{\rm full}} 
\newcommand{\Dack}{D_{\rm ack}}

\usepackage{caption}
\usepackage{subcaption}
\begin{document}

\title{Joint Retransmission, Compression and Channel Coding for Data Fidelity under Energy Constraints}

\author{\IEEEauthorblockN{Chiara~Pielli,~\IEEEmembership{Student~Member,~IEEE,} \v{C}edomir~Stefanovi\'c,~\IEEEmembership{Senior Member,~IEEE,}\\
Petar~Popovski,~\IEEEmembership{Fellow,~IEEE,}
and~Michele~Zorzi,~\IEEEmembership{Fellow,~IEEE}
\thanks{Chiara Pielli (piellich@dei.unipd.it) and Michele Zorzi (zorzi@dei.unipd.it) are with the Dept. of Information Engineering, University of Padova, Italy.}
\thanks{\v{C}edomir~Stefanovi\'c (cs@es.aau.dk) and Petar Popovski (petarp@es.aau.dk) are with the Dept. of Electronic Systems, Aalborg University, Denmark.}
\thanks{A preliminary version of this paper has been presented at IEEE SECON 2017~\cite{Pielli-2017}.}
}}

\maketitle

\begin{abstract}
We consider a monitoring application where sensors periodically report data to a common receiver in a time division multiplex fashion.
The sensors are constrained by the limited and unpredictable energy availability provided by Energy Harvesting (EH), and by the channel impairments.
To maximize the quality of the reported data, the packets transmitted contain newly generated data blocks together with up to $r - 1$ previously unsuccessfully delivered ones, where $r$ is a design parameter; such blocks are compressed, concatenated and encoded with a channel code.
The scheme applies lossy compression, such that the fidelity of the individual blocks is traded with the reliability provided by the channel code.
We show that the proposed strategy outperforms the one in which retransmissions are not allowed.
We also investigate the tradeoff between the value of $r$, the compression and coding rates, under the constraints of the energy availability, and, once $r$ has been decided, use a Markov Decision Process (MDP) to optimize the compression/coding rates.
Finally, we implement a reinforcement learning algorithm, through which devices can learn the optimal transmission policy without knowing a priori the statistics of the EH process, and show that it indeed reaches the performance obtained via MDP.
\end{abstract}

\section{Introduction} \label{sec:intro}

\IEEEPARstart{I}{n} the last few years, Wireless Sensor Networks (WSNs) and their evolution into the Internet of Things (IoT) have spurred some significant research efforts~\cite{Zorzi-2010}.
A major challenge in such IoT applications is to ensure uninterrupted service with minimal device maintenance, driven by the need to minimize the operational expenses.
A typical periodically-reporting device is battery-driven, but expected to operate for long periods without human intervention: in particular, the industry aims to achieve a minimum of 10 years of battery lifetime~\cite{nokia, ericsson, huawei}. 

Energy harvesting is a promising technique through which sensors can scavenge energy from the environment and replenish their batteries, thereby fostering self-sustainability of the devices and, thus, of the IoT application.
Ideally, EH could guarantee infinite lifetime; however, its intermittent nature requires the use of flexible protocols able to adapt to a stochastic energy availability.
A general overview of recent advances in wireless communications with EH is presented in \cite{Uetal2015}, while \cite{Gunduz-2014} discusses the challenges of designing an intelligent EH communication system and presents a general mathematical model that can be adapted to some specific contexts.

In this work, we study how to combine an efficient energy utilization with Quality of Service (QoS) requirements, in terms of quality of the reported data.
We consider a monitoring system where multiple sensor nodes report their readings to a common receiver in a single-hop access network.
Given the predictability of the reporting patterns, the devices access the channel according to a Time-Division Multiple Access (TDMA) scheme.
Each device aims at using its available energy in such a way to optimize the reconstruction fidelity, which depends on the distortion introduced by lossy compression and on the channel impairments.
To this end, we develop a joint source-channel coding mechanism, which comprises transmission of new data, possible retransmission of previously unsuccessfully received data, and lossy compression.
Specifically, each transmitted packet contains a newly generated data block and up to $r - 1$ previous data blocks that were not successfully received, all of them compressed so as to fit the fixed frame size; the retransmission mechanism will be thoroughly explained in Section~\ref{sec:retx}.
The compression is lossy and affects data quality, but on the other hand reduces the volume of information bits to send over the channel, allowing the use of more redundancy to combat the channel impairments.
The selection of the optimal operating strategy in terms of the tradeoff between selection of compression rate and channel coding rate is constrained by the energy availability and statistics of the EH process, and, at the same time, driven by the target minimization of the distortion of the reported data.
In the paper, we investigate the impact of the value of $r$ 
on the distortion, and once $r$ has been decided, use a MDP to study the tradeoff between the compression and the channel coding rates. MDPs are often employed to derived energy management policies, as they represent an appealing solution to optimize some long-term utilities in the presence of stochastic EH~\cite{tutorial}.


Moreover, in the paper we implement a Reinforcement Learning (RL) algorithm, namely R-learning~\cite{schwartz-1993}, through which a device can learn the optimal transmission policy through a trial-and-error discovery process and, thus, overcome the need to know the statistics of the network and the energy dynamics in advance. 
In particular, RL algorithms are commonly used to solve problems that are modeled as MDPs~\cite{Sutton-98} and several works in the literature use them in energy management problems in the presence of EH, cf. \cite{Blasco-2013, Ortiz-2016}.
We also note that more sophisticated approaches, like deep learning algorithms, cannot be pursued because they have too demanding resource and computation requirements that the simple IoT devices could not bear.

The key contributions of this work are summarized as follows:
\begin{itemize}
 \item We set up a transmission problem where energy consumption and data quality are balanced. We consider both lossy compression and the impairments due to transmission over a fading channel, and propose an exhaustive parametric model of the energy dynamics of a device.
 
 \item We jointly consider source and channel coding, accounting for realistic rate-distortion curves that match those of practical data compression algorithms, and characterize the outage probability that affects the transmission of short packets over a fading channel. 
 
 \item Driven by the convexity of the distortion function, we introduce a hybrid retransmission mechanism where previously lost packets are sent along with the new one. Although this implies choosing a larger compression ratio so that all packets can fit in the same slot, the long-term average distortion is improved. Further, by fixing the maximum number of retransmissions allowed, latency is limited.
 
 \item We implemented an RL algorithm that learns from experience by trial-and-error and gradually converges to the optimal policy determined through the MDP.
 
 \item We evaluate numerically the improvement brought by the retransmission scheme, as well as 
the role that various system variables play in determining the performance.
\end{itemize}
A preliminary version of the work presented in this paper, focused on the MDP solution to the problem of joint selection of compression and channel-coding ratio, appeared in \cite{Pielli-2017}.
This paper presents its substantial extension, where the retransmission mechanism and the RL algorithm constitute the two main novel contributions.

The rest of the paper is structured as follows.
We conclude this section with an account of the related work.
Section~\ref{sec:comm_model} describes the communication model, i.e., how data is processed and transmitted, while Section~\ref{sec:en_model} focuses on the energy consumption and EH profiles. In Section~\ref{sec:retx} we present our energy- and QoS-aware retransmission scheme. The optimization problem is defined in Section~\ref{sec:problem} and optimally solved in Section~\ref{sec:solution}; Section~\ref{sec:results} shows the numerical evaluation. In Section~\ref{sec:learning} we discuss how nodes can learn the optimal policy. 
Finally, the conclusions are drawn in Section~\ref{sec:conclu}.

\subsection*{Related work}
Investigating the effects of packet losses on data distortion is not a new topic.
However, many works limit their studies to Gaussian data sources or neglect the energy limitations, cf.~\cite{Zachariadis, Aguerri-2011}.
A common approach is to use the distortion exponent as the performance metric~\cite{Laneman,Zhao-2016}, but this is meaningful only for the high Signal-to-Noise Ratio (SNR) regime.
Since the SNR is generally low in IoT scenarios, and especially in EH-powered nodes, we do not make use of the distortion exponent, but rather consider a distortion metric that accounts for compression and channel outages and that is affected also by the energy availability.
Another beaten path is that of layered transmission schemes, where the source is coded in superimposed layers, like in~\cite{etemadi}.
Each layer successively refines the data description and is transmitted with a larger coding rate, thus the transmission is less robust to failures.
This practice is often used in multimedia applications~\cite{ng-2014}, but is not very meaningful in other contexts. 

Several works focus on the minimization of data distortion in the presence of energy limitations and/or EH.
For example, \cite{bhat} analyses the tradeoff between the energy required by quantization and transmission in the presence of EH, but neglects the effect of packet losses on the data quality at the receiver.
A similar problem is treated in~\cite{Castiglione}, where the sensor nodes jointly perform source-channel coding and balance energy between processing and transmission in such a way to guarantee a minimum average distortion and, at the same time, maintain the data queue stable.
Also~\cite{motlagh} proposes a joint source-channel coding scheme for Gaussian and binary sources.
The authors derive lower bounds on the distortion achievable when the energy buffer may have some leakage.
In~\cite{bui}, the sensor nodes can tune their duty cycle and information generation rate with the objective of guaranteeing energy self-sufficiency to a multi-hop network.

Two works that bear similarities with ours are~\cite{zordan} and~\cite{Fullana-2017}.
In~\cite{zordan}, the goal is to maximize the long-term average quality of the transmitted packets by adapting  the degree of lossy compression and the transmission power. Power control is used to maintain the packet error probability below a chosen threshold, depending on the state of the channel.
Like in our work, the optimal transmission strategy is determined through a MDP. 
In~\cite{Fullana-2017}, the reconstruction of time-correlated sources in a point-to-point communication is formulated as a convex optimization problem and solved with an iterative algorithm.
The node has to decide on the transmission power and rate and is subject to EH constraints.

It is difficult to find works that deal with retransmissions in conjunction with data processing outside of the multimedia networks.
Retransmissions in IoT constrained networks are considered, e.g., in~\cite{aprem-2013}, which addresses the problem of transmit power control in the presence of EH when a sensor node makes use of an Automatic Repeat Request (ARQ) protocol and retransmits the lost packets. Differently from us, they assume that some Channel State Information (CSI) is available at the nodes, and, further, they do not consider data processing.
In~\cite{Costa-2013}, sensor nodes implement an energy-aware hop-by-hop retransmission mechanism where only packets carrying critical information are ensured to be retransmitted. This differs from our scheme, where packets are not differentiated according to priorities but can be compressed at the source.

The retransmissions are also often investigated in terms of cooperative retransmissions, where devices collaborate with each other, cf.~\cite{Hu-2013,Lee-2012}.
However, in this work we do not to consider cooperation among nodes, as we are interested in scenarios where the devices are unaware of the presence of the others and of the resources they have available, and communicate solely with the gateway, e.g., as in a LoRa network~\cite{lorawan_specs}.



\section{Communication model} \label{sec:comm_model}
 
\begin{figure*}[t]
	\centering
	\includegraphics[width=.8\textwidth]{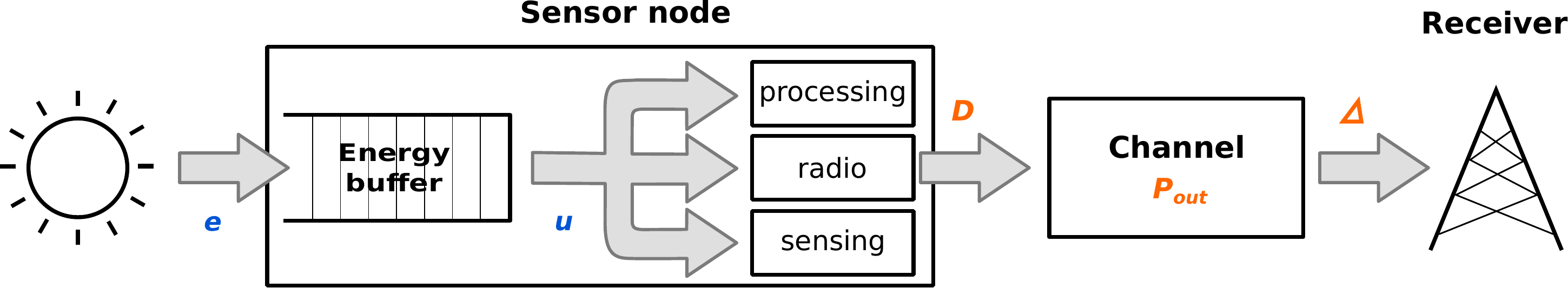}
	\caption{Block diagram of a device. The blue labels indicate the energy flows, while the orange labels are related to the data reporting.}
	\label{fig:node}
\end{figure*}

We assume a \mbox{single-hop}, star topology network, comprising a multitude of IoT devices that periodically monitor some phenomena of interest and report data to a common receiver.
This data collector is assumed to be connected to the energy grid, whereas the sensor nodes are battery-powered, but also endowed with energy harvesting capabilities.

The predictability of the traffic pattern makes TDMA a good choice for transmission scheduling among the devices.
Since in this case the devices do not interfere with each other, we focus just on a single device, with the aim to find its optimal transmission strategy. 
The corresponding model is represented in Fig.~\ref{fig:node}.
A device is endowed with a circuitry to scavenge energy from the environment; this energy is stored in a buffer of finite size and used by the node to generate, process, and send data to the common receiver through a Rayleigh-fading channel.
In this section, we describe the joint source-channel coding scheme, while in Section~\ref{sec:en_model} we discuss the energy harvesting aspects.
For the sake of simplicity, we focus on the case when no retransmissions are allowed, while the extension that takes into account retransmissions is discussed in Section~\ref{sec:retx}.

\subsection{Compression at the source}
 
A device is capable of compressing in a lossy fashion the time series generated through sensing the environment, as described next.
Between two consecutive reporting events, a sensor node collects a block of readings\footnote{We will refer to blocks of readings as data blocks, to distinguish them from the packets sent over the communication channel after processing}, where each block has a constant size of $L_0$ bits and is independent of the previous ones. 
The device compresses the generated block by selecting a compression level $k$, where $k\in\{0,\dots,m\}$, and produces a compressed block of size $L$ bits.
When $k=0$, the size of the compressed block is $L=0$, i.e., no packet is transmitted, and when $k=m$, the size of the compressed block is $L = L_0$, i.e., no compression takes place.
We define the \emph{compression ratio} as the ratio between the size of the compressed block and that of the original one: $L/L_0\!=\!k/m\!\in\![0,1]$. 
Lossy compression makes it possible to trade some accuracy in the data representation using a lower compression ratio with additional error-correction redundancy, and consequently have an increased robustness to channel impairments, as will be explained in Section~\ref{ssec:data_tx}. 
The distortion metric we employ is defined as the maximum absolute error between the original and the compressed signal normalized to the amplitude range of the signal in the considered time window; it has been used on real data series in, e.g.,~\cite{Zordan-2017}.


The \mbox{compression ratio-distortion} curve is signal- and algorithm-dependent. 
We consider an automatic sensor profiling approach where a device dynamically decides upon the compression algorithm to use depending on the type of signal it generates and the corresponding distortion.
We use the following expression for the distortion introduced by lossy compression, that was derived in our previous work~\cite{eccentric}:~ 
\begin{equation} \label{eq:dist}
D(k) = \begin{cases}
        b \left( \left(\frac{k}{m} \right)^{-a} - 1\right) \, &\text{if } k\ge 1 \\
        D_\fl \, &\text{if } k = 0
       \end{cases}
\end{equation}
where $b\!>\!0$, $0\!<a\!<\!1$.
The choice $k=0$ entails that the packet is discarded (due to energy restrictions) and the corresponding distortion is equal to the maximum value $D_\fl \triangleq 1$. 
Eq.~\eqref{eq:dist} shows that the distortion is a convex and decreasing function of the compression ratio $k/m$.


\subsection{Data transmission} \label{ssec:data_tx}

Each IoT node transmits during its dedicated time slot, whose  duration $T$ defines the maximum number of bits that can be sent $S=T/T_{\rm b}$, with $T_{\rm b}$ being the (fixed) bit duration. It is reasonable to assume that $L_0\le S$, i.e., a device may avoid compressing a packet. 
After compression, there are $L \le L_0$ information bits and the corresponding coding rate is $R\!=\!L/S$.
Depending on $R$ and the actual channel conditions, the packet may not be correctly received with an outage probability $\pr_\out(R)$.

We envisage the presence of an acknowledgment mechanism, so that the receiver sends feedback to the transmitter; however, no channel state estimation is performed, and 
consequently the transmission power $P_\ttx$ is kept constant.
The communication channel is affected by block Rayleigh fading; when the channel is in a deep fade the packet is lost and an outage occurs.

Because the packets sent by IoT devices are likely to be short, we exploit the recent results of finite-length information theory that adapt the classical concepts of channel capacity to the case of short data packets~\cite{polyansky}. In particular, the results of~\cite{yang} legitimate the use of the quantity $\log_2(1+\gamma)$ to represent the maximum rate even in the finite-length regime, where $\gamma$ is the SNR at the receiver, given by:~
%
\begin{equation}
  \gamma = \dfrac{|H|^2 P_\ttx}{A^2\, (d/d_0)^{\eta} N} \triangleq |H|^2 \, \bar{\gamma}.
\end{equation}
$H$ is the channel gain coefficient that represents fading (assumed to be constant over the packet duration in the quasi-static scenario), and $\bar{\gamma}$ is the expected SNR at the receiver, that depends on the transmission power $P_\ttx$, the noise power $N$, and the term $A^2\, (d/d_0)^{\eta}$ that accounts for path loss.
The latter depends on the path-loss exponent $\eta$, the distance between transmitter and receiver $d$, and a path-loss coefficient $A=4\pi d_0 f_0/c$, where $f_0$ is the transmission frequency, $c$ the speed of light, and $d_0$ a reference distance for the antenna far field~\cite{Rappaport-1996}.
Thus, the outage probability can be approximated as:~
\begin{equation}
 \pr_\out(R) = \prr\left(\,\log_2\left(1+\gamma\right) < R\right). \label{eq:out_gen}
\end{equation}

As we consider Rayleigh fading, $H$ follows a complex Gaussian distribution with zero mean and unit variance.
In this case, the outage probability becomes:~
\begin{equation} 
 \pr_\out(R) = 1 - e^{-(2^{R}-1)/\bar{\gamma}}, \label{eq:outage}
\end{equation}
which is non-decreasing in $R$ (and thus in $k$, since $L=k/m \,L_0$), and initially convex and then concave.
Clearly, the farther the device from the receiver, the larger the outage probability, since $\bar{\gamma}$ decreases with distance.
If a packet gets lost, the block of readings contained in it can be compressed and transmitted again along with the subsequently generated blocks, as described in Section~\ref{sec:retx}.
After $r$ failed transmission attempts, the block is considered outdated and discarded.

\section{Model of energy dynamics} \label{sec:en_model}

The energy dynamics of the devices strongly influence the system performance, and both data processing and transmission policies should dynamically adapt to the resource availability. Designing such an energy-aware framework requires to accurately model the energy inflow and consumption.

\subsection{Energy consumption} \label{ssec:energy_cons}

In the following, we describe a parameterized model that tries to capture all the major sources of energy expenditure: communication, data acquisition, processing, and circuitry.
We are not interested in the energy spent for packet reception because the data collector does not have energy constraints.

\textbf{\textit{Data processing.}} 
In \cite{Zordan-2014}, the energy consumed by processing algorithms is evaluated by mapping the number and type of arithmetic instructions into the corresponding energy drain:~ 
 \begin{align} \label{eq:e_processing}
   E_p(k)  = 
   \begin{cases}              
      E_0 \, L_0 \, N_p(k) & \text{if }  1\le k \le m-1\\
      0 & \text{if } k=0,m
   \end{cases}  
 \end{align}
where $E_0$ is the energy consumption per CPU cycle (which depends on the \mbox{micro-controller} unit), and $N_p(k)$ is the number of clock cycles required by the compression algorithm per uncompressed bit of the input signal (and depends on the compression ratio). If the packet is not compressed ($k=m$) or is discarded ($k=0$), no energy is consumed.

We assume that the devices employ the Lightweight Temporal Compression (LTC) algorithm, which is a widely used lightweight compression techniques for wireless sensors.
In this case, the function $N_p(k)$ is increasing and concave in $k$~\cite{Zordan-2014}:~
\begin{equation}
N_p(k) = \alpha_p \frac{k}{m} + \beta_p, \quad 1 \le k \le m-1 ,
\end{equation}
with $\alpha_p, \beta_p > 0$. Notice that the more compressed the packet, the less the energy spent.
This seemingly counterintuitive fact is due to implementation details and we refer the reader to~\cite{Zordan-2014} for explanation.
Finally, we do not account for the contribution of channel encoding, because it is typically negligible~\cite{Howard-2006}.

\textbf{\textit{Transmission.}}
 The energy cost of a wireless transmission with power $P_\ttx$ for a period of length $T$ can be modeled as:~
 \begin{equation}
   E_{{\rm tx}}  =  \frac{T\, P_{{\rm tx}}}{\eta_{\rm A}}, \label{eq:e_tx}
 \end{equation}
 \noindent where $\eta_{\rm A}\in (0,1]$ is a constant that models the efficiency of the antenna's power amplifier.

\textbf{\textit{Sensing and circuitry.}}
We assume that the periodical sensing drains a constant amount of energy $\beta_s$ in the window between two transmission slots.
Also the energy required for switching between idle and active mode and maintaining synchronization with the receiver can be represented with a constant term $\beta_c$. 
Finally, we need to account for the additional energy spent by the circuitry during a transmission, $\mathcal{E}_c \,T$, where $\mathcal{E}_c$ is a circuitry power. Hence:~
\begin{align}
 E_c(k) =  \beta_s + \beta_c + \mathcal{E}_c \,T\cdot \chi_{\{k>0\}},  \label{eq:e_circuitry}
\end{align}
where $\chi_{\{k>0\}}$ is the indicator function equal to 1 if $k\!>\!0$ and to zero otherwise (recall that the packet is dropped if $k=0$).

\medskip



\subsection{Energy harvesting and battery dynamics} \label{ssec:eh}
%
%

The sensor nodes are not connected to the energy grid, but are provided with some energy-scavenging circuitry and can collect energy from the environment.
We assume that the energy supply is time-correlated (e.g., solar power); for this case, there exist models in the literature that have been built and validated against real data.
In particular, both~\cite{Ku} and \cite{solarstat} proved that a time-correlated energy supply can be accurately described through a stochastic Markov process, where each possible state entails a different distribution of the energy income.
When dealing with energy harvesting, the model is generally discrete, i.e., the energy inflow is quantized.

Accordingly, we track the dynamics of the source through an $X$-state Markov Chain (MC): the source is in state $x\in\mathcal{X}=\{0,\dots, X-1\}$ and scavenges $e\in\{0,\dots,E\}$ quanta of energy from the environment, according to some probability mass function. 
Our framework is general and can accomodate any assumptions about the number of states of the harvesting process and the harvesting statistics in each state.
However, to obtain some representative results, in this paper we consider a 2-state MC ($X=2$).  
In particular, $x=0$ represents a low energy state (e.g., night) during which no energy can be harvested ($e=0$); when $x=1$, the source is in a high energy state (e.g., day) and the energy income follows a truncated discrete normal distribution, $e \sim \mathcal{N}(\mu, \sigma^2)$ in the discrete interval $\{1,\dots\,E\}$ (analogous to~\cite{zordan} and~\cite{Ku}).

We assume that the battery of a node has a finite size $B$.
The temporal evolution of the battery can be modeled as:~
\begin{equation} \label{eq:battery}
 b' = \min\left\{b + e -  u, B\right\} \triangleq \left(b + e -  u\right)^\dagger,
\end{equation}
where $b$ and $b'$ respectively represent the current and next battery level, and $u$ is the energy used in the current slot, which depends on processing, transmission and circuitry, as given in Eqs.~\eqref{eq:e_processing}, \eqref{eq:e_tx} and \eqref{eq:e_circuitry}.
Due to the energy causality principle, it is:~
\begin{equation} \label{eq:en_causality}
 u \le b.
\end{equation}

An overly aggressive or conservative energy management could either deplete the battery ($b=0$) or fail to use the excess energy and waste it ($b=B$).
These situations need to be prevented by designing a scheme that dynamically adapts to the randomness of the energy inflow, so as to ensure acceptable performance on a long-term horizon.

\section{Energy-aware (re)transmission scheme} \label{sec:retx}

Here, we extend the joint source-channel coding scheme described in Section~\ref{sec:comm_model} to account for packet retransmissions. 


\begin{figure}[t]
	\centering
	\includegraphics[width=\columnwidth]{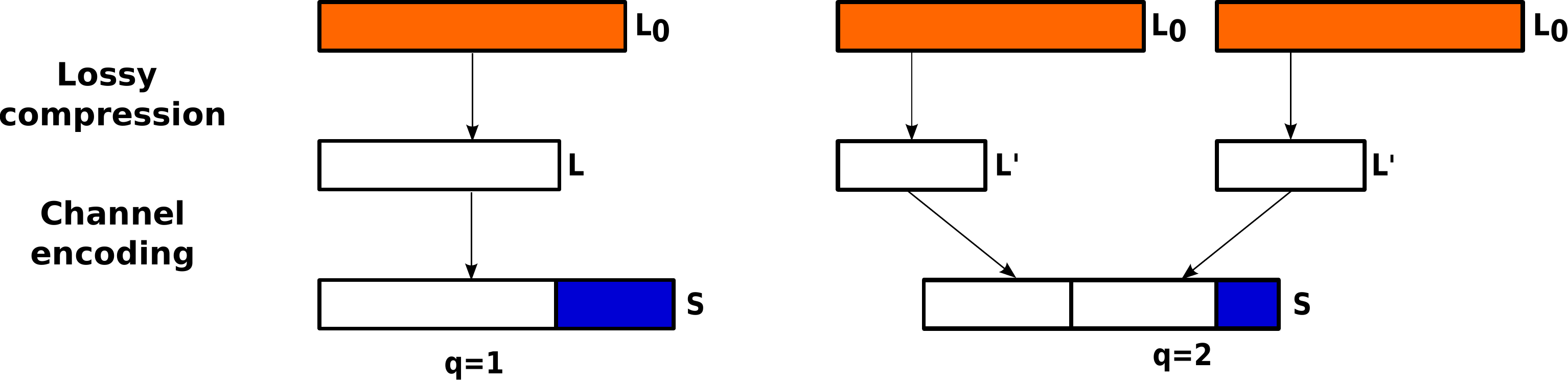}
	\caption{The retransmission mechanism: encoding procedure.}
	\label{fig:retx}
\end{figure}

The acknowledgment feedback allows the transmitter (i.e., the reporting device) to know whether a packet has been correctly received at its intended destination.
When an outage occurs, the packet is lost and needs to be retransmitted.
However, if a simple ARQ protocol is employed, the delay for sending newly generated data blocks will increase, depending on the number of the devices participating in the TDMA scheduling as well as on the number of allowed retransmissions, and can potentially become unbounded.
Consequently, it may happen that the newly generated data blocks will become outdated, without having a chance to be transmitted.
To prevent this, we consider a retransmission scheme in which nodes send their new data blocks together with the previously lost data blocks (if any) in the same time slot, as explained further.

Suppose that the transmission of a data block fails; in the successive time slot, the device will try to send the new block together with the previously lost one.
Thus, the device has to compress and transmit two blocks of original size $L_0$ each within the $S$ channel uses available.
If this transmission also fails, in the next slot the sensor node will process and transmit the new block and the two that were previously lost.
In general, if the last $q$ transmissions failed, the sensor node has to process $q+1$ blocks of $L_0$ bits each and send them together. 
We assume that a maximum number $r$ of transmission attempts can be made for each piece of data, where the value of $r$ is dictated by the application, e.g., because of latency or QoS considerations.

Recall that data blocks are of a fixed size $L_0$ and are assumed to be independent of each other.
Because of the independence assumption, when multiple data blocks are sent in the same time slot, the information they contain is not fused or processed jointly, but the compression is done separately for each of them.
For the sake of fairness, we treat the data blocks in the same way, i.e., we apply to them the same compression ratios and obtain $q$ compressed blocks of size $L$, where $q = r + 1 $.
These are then joined in a single block of size $q\,L$, which is then encoded, producing a packet of size $S$, and transmitted.
Thus, the same distortion is introduced at the source for all data blocks placed in the same packet, but then they are treated as a single entity which is sent over the channel and subject to a certain outage probability that depends on the coding rate $R=q L/S$.
Fig.~\ref{fig:retx} shows the encoding mechanism when a single block (on the left) or two blocks (on the right) are transmitted in the same slot. 

%


\begin{figure}[t]
	\centering
	\includegraphics[width=.9\columnwidth]{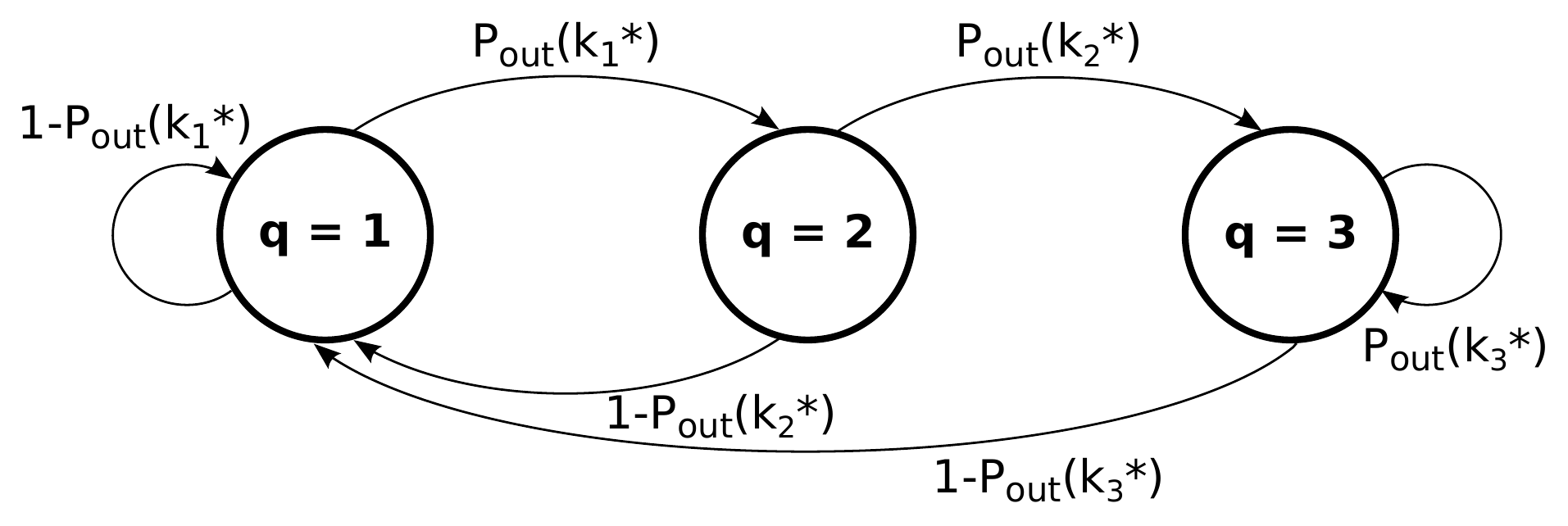}
	\caption{Structure of the MC that models the dynamics of the backlog state $q$ for $r=3$.}
	\label{fig:queue}
\end{figure}

Fig.~\ref{fig:queue} shows the evolution of the data queue, which behaves like a success-runs MC.
If the data queue in a certain slot has size $q$ and the transmitted packet is lost, the queue state becomes $q' = \min\{q+1, r\}$, otherwise the new queue state is $q'\!=\!1$.
We denote as $\Delta_q(k)$ the distortion at the receiver when the queue state is $q$ and the chosen compression ratio is $k$. Such distortion depends on the reception outcome: in case of failure (NACK), the distortion at the receiver is set to $\Delta_q(k) = D_\fl$, even though the data blocks could be retransmitted; instead in case of successful transmission (ACK), we sum all the distortions of the $q$ blocks that have been compressed with the same ratio $k/m$, but we also subtract all the penalties that were obtained for the previous packet losses we accounted for, since the blocks were eventually successful. Thus, in this latter case, it is $\Delta_q(k) = \Dack(q,k)$ with:~
\begin{equation}
 \Dack(q,k) =  q\,D(k) - (q-1) D_\fl. \label{eq:dack} 
\end{equation}

We denote as $k_q^\star$ the optimal value of $k$ (i.e., the one that determines the optimal source-channel coding scheme to use) when $q$ packets share the same time slot; in Section~\ref{ssec:RDP} we explain how to determine it.

\smallskip

\textit{\textbf{Energy consumption.}}
\noindent Suppose that the data queue state is $q$.
Between two consecutive slots, the node consumes energy to process and transmit all $q$ data blocks.
However, the energy required by transmission and circuitry does not depend on $q$, because the duration of the transmission is always $T$, i.e., $S$ bits, see Eqs.~\eqref{eq:e_tx} and~\eqref{eq:e_circuitry}.
Hence, the energy consumed by the node when there are $q$ data blocks in the queue and it uses a compression ratio equal to $k/m$ is:~
\begin{equation} \label{eq:energy_used}
  q\,E_p(k) + E_\ttx\cdot\chi_{\{k>0\}} + E_c(k).
\end{equation}
The optimization described in Section~\ref{ssec:RDP} determines the optimal source-coding scheme, i.e., the value $k_q^\star$, and the corresponding energy consumption is obtained by substituting $k=k^\star_q$ in Eq.~\eqref{eq:energy_used}. 

 
\section{Problem formulation} \label{sec:problem} 
 
Here, we mathematically define the objective of our optimization problem and describe the structure of the MDP, while the solution technique is explained in Section~\ref{sec:solution}.

\subsection{The optimization objective} \label{ssec:objective}
Our objective is to maintain for each node an energy-neutral operation mode while minimizing the long-term average distortion at the receiver, which depends on the outcomes of the transmissions, as explained in the previous section.
If we consider $q$ data blocks of size $L_0$ that are compressed with the same compression ratio $k/m$ and encoded jointly at rate $R = qL(k)/S$, the expected distortion at the receiver is:~
\begin{align}
 \mathbb{E}[\Delta_q(k)] = & \; qD(k)\,\left(1- \pr_\out\left(q\frac{L(k)}{S}\right)\right) \nonumber\\
 +& \;q D_\fl\,\pr_\out\left(q\frac{L(k)}{S}\right), \label{eq:delta}
\end{align}
where $\pr_\out(\cdot)$ is the outage probability as given in Eq.~\eqref{eq:out_gen}, $qD(k)$ and $q D_\fl$ are the distortions obtained when the packet is acknowledged or lost, respectively.

In other words, if the packet is successfully received, its distortion corresponds to that introduced at the source for all initial $q$ blocks of measurements, otherwise we account for the maximum distortion level for all packets, as if they had not even been sent.
We are interested in the expected distortion at the receiver, thus we weight the two cases with their probabilities.
Notice that the distortion at the source $D(k)$ decreases as $k$ increases, whereas the outage probability decreases for smaller coding ratios, i.e., as $k$ decreases.
This implies a tradeoff between the distortion introduced by the lossy compression and the probability that the transmitted packet will be successfully received, through the choice of the value of $k$. 

To minimize $\mathbb{E}[\Delta_q(k)]$ and guarantee self-sufficiency of the network, it is necessary to (i) decide on $k$, and (ii) in each slot, allocate the energy consumption based on the current battery level, the dynamics of the energy source, and the energy consumption profile, in such a way to prevent energy outages (that disrupt the communication) and battery overflows (that waste energy). 
We formulate the problem by means of an MDP, 
where the actions correspond to the energy to use 
operations while the costs are represented by the expected distortion at the receiver.
By doing so, the energy self-sufficiency of the node is ensured and the QoS is optimized.

 
 \subsection{The Markov Decision Process} \label{ssec:MDP}
The MDP is defined by the tuple $\big(\mathcal{S},\,\mathcal{U},\,P,\,c(\cdot)\big)$, where $\mathcal{S}$ denotes the system state space, $\mathcal{U}$ is the action set space, $P$ is the set of transition probabilities of the system state space and $c(\cdot)$ is the associated cost function for taking an action.

\textbf{\textit{System state space}} $\mathcal{S} \triangleq \mathcal{X} \times \mathcal{B} \times \mathcal{Q}$, where $\mathcal{X}=\{0,1\}$ represents the set of energy source states, $\mathcal{B}=\{0,\dots,B\}$ the set of energy buffer states, and $\mathcal{Q}=\{1,\dots,r\}$ the set of backlog states, i.e., the number of waiting packets.
Notice that we need to keep track of the data queue size, unlike in our previous work~\cite{Pielli-2017} where it was always one.

\textbf{\textit{Action set space}} $\mathcal{U} \triangleq \{0,\dots, B\}$.
In each slot, the device observes the current system state $s\in\mathcal{S}$ and decides how much energy $u \in \mathcal{U}_{s} \subseteq \mathcal{U}$ to use to process and transmit the data it collected.
In accordance with~\eqref{eq:en_causality}, this quantity cannot exceed the battery level, i.e., $\mathcal{U}_{s} = \{0,\dots,b\}$. 

\textbf{\textit{Transition probabilities}} $P$ govern the system dynamics.
The probability of going from state $s = (x,\, b,\, q)$ to $s' = (x',\, b',\, q')$ with action $u$ is:~
 \begin{equation}
  \begin{split}
   \prr(s'|s, u)=   &\; \; \: p_x (x'|x) \cdot p_e (e|x)\cdot p_q (q'|q,u)\\ 
   & \cdot \, \delta \left(b' - \left(b + e - u \right)^\dagger \right)
  \end{split}
 \end{equation}
where $p_x (x'|x)$ is obtained from the transition probability matrix of the MC that models the source state,  $p_e (e|x)$ is the mass distribution function of the energy inflow in state $x$ (see Section~\ref{ssec:eh}), and $p_q (q'|q,u)$ represents the probability that the backlog size goes from $q$ to $q'$ when action $u$ is taken. The last term $\delta(\cdot)$ is equal to $1$ if its argument is zero, and zero otherwise, and ensures that the transitions between states are consistent with the dynamics of the battery level, see Eq.~\eqref{eq:battery}.

\textbf{\textit{Cost function}} $c(\cdot)$. When the sensor node is in state $s=(x,b,q)$ and selects action $u$, it implicitly decides upon the source-channel coding scheme that minimizes $\mathbb{E}[\Delta_q(k)]$, i.e., it decides the optimal number $k^\star_{q}$ of information bits to send over the channel per data block, given the available energy $u$.
The transition from state $s$ to state $s'$ when action $u$ is taken entails a cost:~
\begin{equation} \label{eq:cost}
 c(s, u, s') = 
 \begin{cases}
  \Dack(q, k^\star_{q}) \; &\text{if } q' = 1\\
  D_\fl \; &\text{otherwise}
 \end{cases} 
\end{equation}
and therefore the cost of choosing action $u$ in state $s$ is:~
\begin{equation}\label{eq:tot_cost}
 \tilde{c}(s, u) = \sum\limits_{s'\in\mathcal{S}} \Pr(s'|s, u) \,c(s, u, s'). 
\end{equation}
Notice that the cost only depends on the data queue component $q$ of the state, while the battery level $b$ affects the set of admissible actions.

When retransmissions are not allowed ($r=1$), the backlog state is always $q\!=\!q'\!=\!1$, whatever the outcome of the transmission is. In this case, the MC of Fig.~\ref{fig:queue} reduces to a single state, and Eq.~\eqref{eq:cost} is no longer meaningful, thus when $r=1$ we set $c(s, u, s') \equiv \tilde{c}(s, u) \equiv \mathbb{E}[\Delta_1(k^\star_1)]$.

\section{Optimal policy} \label{sec:solution}

The optimal policy is the one that, based on the statistics of the energy harvesting process and the current battery level, decides how much energy to use in order to guarantee the lowest average distortion at the receiver.
We first discuss how to determine the optimal source-channel coding scheme, hence $k^\star_q$ for each possible backlog state $q\le r$, and then describe how to solve the MDP optimally.

\subsection{Rate-distortion tradeoff} \label{ssec:RDP}

\begin{figure}
 \centering
 \setlength\fwidth{.85\columnwidth}
 \setlength\fheight{0.5\columnwidth}
 \input{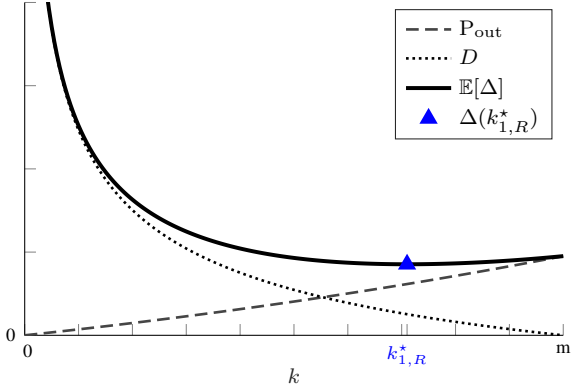}
 \caption{Example of location of $k^\star_R$, which is the point of minimum of Eq.~\eqref{eq:delta}, for $q=1$.}
 \label{fig:k_r}
\end{figure}

In~\cite{Pielli-2017}, we proved that for the case $q=1$ the expected distortion at the receiver given in Eq.~\eqref{eq:delta} exhibits a unique point of minimum $k_{1,R}^\star$ when $D(k)$ and $P_\out(k)$ are characterized as in Eqs.~\eqref{eq:dist} and~\eqref{eq:outage}, respectively.
An example of this is shown in Fig.~\ref{fig:k_r}.
This is still valid when retransmissions are introduced, as can be inferred by looking at Eq.~\eqref{eq:delta}. 
The $q$ packets are compressed separately but with the same compression ratio, which cannot exceed $m/q$, hence $D(k)$ is truncated for $q>1$.
Instead, the outage probability maintains the same shape since the packets are encoded together.
Hence, there exists an optimal point $k_{q,R}^\star$, whose value depends on the number of packets $q$ that are sent in the same time slot and that minimizes the expected distortion at the receiver.
If the device uses $k < k_{q,R}^\star$, the packet will go through the channel with a higher probability but its distortion will be larger; on the other hand, if $k > k_{q,R}^\star$, the distortion will be smaller, but it is more likely that the packet will be lost.

If the amount of energy allocated $u$ allows it, the device will choose the optimal coding scheme corresponding to $k^\star_{q,R}$, otherwise it simply selects the maximum possible $k$ dictated by the energy constraint, because $\mathbb{E}[\Delta_q(k)]$ is decreasing if $k\le k_{q,R}^\star$, which is due to how $k_{q,R}^\star$ is defined.
The energy consumption, see Eq.~\eqref{eq:energy_used}, is non-increasing in $k$, and we denote as $k_{q,E}^\star(u)$ the largest value of $k$ that solves $q\,E_p(k) + E_\ttx\cdot\chi_{\{k>0\}} + E_c(k) \le u$ for a given $q$, see Eq.~\eqref{eq:energy_used}.
Then, when the backlog state is $q$, the device will choose the source-channel coding scheme corresponding to:~
\begin{equation} \label{eq:k_star}
 k^\star_{q} = \min\{k_{q,R}^\star, k_{q,E}^\star(u)\}.
\end{equation}
Such value clearly depends on the energy $u$ that the node decides to employ, but we omit denoting this explicitly in favor of a lighter notation.
In~\cite{Pielli-2017} we showed that the expected distortion at the receiver is convex for $k\le k_{q,R}^\star$ and, consequently, is a convex non-increasing function of $u$. 
 
The choice of how much energy to use when in state $s=(x,b,q)$ uniquely determines the joint source-channel coding scheme (i.e., the number of transmitted information bits per data block) that leads to the smallest expected distortion at the receiver.
For further details, we refer the reader to~\cite{Pielli-2017}.



\subsection{Solving the MDP} \label{ssec:RVI}

When the sensor device is in a certain state, it selects the action to take according to a policy $\pi:\mathcal{S}\rightarrow \mathcal{U}$.
The corresponding long-term average cost is:~
\begin{equation} \label{eq:J}
 J^\pi(s) = \lim\limits_{M\rightarrow +\infty} \frac{1}{M} \mathbb{E}_s\left[ \sum\limits_{m=0}^{M-1} \tilde{c}(s_m, u_m) \middle\rvert s_0 = s \right],
\end{equation}
where the initial state $s_0$ is given.
Notice that each decision affects all subsequent decisions.

We want to determine the optimal policy $\pi^\star$, i.e., the set of rules that maps each system state into the optimal action with respect to the average cost criterion.
The MDP defined in Section~\ref{ssec:MDP} has unichain structure and bounded costs, implying that Eq.~\eqref{eq:J} does not depend on the initial state:~
\begin{equation}
 J^\pi(s)\equiv J^\pi, \quad \forall s\in\mathcal{S}.
\end{equation}
Hence, we can restrict our search to 
Markov policies only~\cite{altman}.

 
To determine $\pi^\star$, we used the Relative Value Iteration Algorithm (RVIA), which is a variant of the well-known Value Iteration algorithm for average long-term problems and provably converges~\cite{bertsekas}.
To understand why it works, we first define the $n$-step value-function by induction as:~
\begin{equation} \label{eq:v}
 v_{n}(s) = \min_{u\in\mathcal{U}_{s}} \left\{ \tilde{c}(s,u) + \sum\limits_{s'\in\mathcal{S}}\,\Pr(s'|s,u)v_{n-1}(s') \right\},
\end{equation}
where $v_0(s)$ is arbitrarily defined, e.g., $v_0(\cdot)=0$.
Function $v_n(s)$ represents the minimum expected $n$-step cost that can be achieved from an initial state $s$, because it sums the immediate cost $\tilde{c}(s,u)$ obtained in the initial state with the expected optimal cost obtained in the $n-1$ subsequent slots (through $v_{n-1}(\cdot)$ and then recursively).
Based on this, the optimal policy $\pi^\star$ is such that $J^{\pi^\star}(s) = \lim_{n\rightarrow\infty} v_n(s)/n$, and as $n$ grows the dependence on $s$ fades ($J^{\pi^\star}(s) \equiv J^{\pi^\star}$).


RVIA leverages on this and defines two functions $J$ and $Q$ that are alternatively updated starting from an initial estimate $J_0(\cdot)$ until convergence:~
\begin{align}
  & Q_{j}(s,u) = \tilde{c}(s,u) + \sum\limits_{s^\prime\in\mathcal{S}} \prr(s^\prime|s, u)\,J_{j-1}(s^\prime) \label{eq:Q}\\
  & J_{j}(s) = \min_{u\in \mathcal{U}_s} Q_{j}(s,u), \label{eq:J_Q}
 \end{align}
with $j$ being the iteration index.  
The convergence criterion is chosen as the span seminorm operator $sp(w)\triangleq \max(w)-\min(w)$, because it guarantees that~\eqref{eq:J_Q} is a contraction mapping~\cite{bertsekas}. RVIA is stopped at iteration $l$, when $sp(J_{l+1}(s) - J_l(s)) \le t$ for a chosen threshold $t$.
The optimal policy dictates the action $u^\star(s)$ to take in each state $s$:~
\begin{equation}
u^\star(s) = \argmin_{u\in \mathcal{U}_s} Q_{l}(s,u),
\end{equation}
and entails the following average long-term cost:~
\begin{equation}
 C^\star =\sum\limits_{s\in\mathcal{S}} \rho_s\, \tilde{c}(s,u^\star(s)) \label{eq:final_cost}
\end{equation}
where $\rho_s$ is the steady state probability of state $s$ induced by the optimal policy (the MDP reduces to a MC when the actions to take in each state are deterministic).

\subsection{The effects of retransmitting} \label{ssec:retx_eff}

When a transmission slot is used for multiple packets combined together, these packets will be more compressed and have a larger distortion than when a slot is reserved to a single packet.
However, the convexity of $D(\cdot)$ suggests that combining packets improves the average long-term distortion.
This is formalized in the following result:~
\begin{thm} \label{th:retx}
 For any value $r$ of the maximum number of transmission attempts that can be dedicated to each packet such that $k_{r,R}^\star > 0$, when the energy constraints are neglected, i.e., $k_{r}^\star = k_{r,R}^\star$ and $k_{1}^\star = k_{1,R}^\star$, the retransmission mechanism described in Section~\ref{sec:retx} achieves a lower average long-term distortion compared to the base case where packets can be sent only once.
 \begin{proof}
  See Appendix~\ref{apx:proof_retx}.
 \end{proof}
\end{thm}

Notice that, if $k^\star_r = 0$, there would clearly be no gain in using the retransmission scheme as packets are simply discarded; this is unlikely to happen, especially if $r<m$.

Theorem~\ref{th:retx} proves that the combined retransmission scheme leads to enhanced performance in terms of QoS in the absence of energy constraints. 
We now discuss what happens when these constraints come into play.
The maximum compression ratio allowed by the allocated energy is $k_E^\star(u)/m$, which depends on the energy allocation $u$ and on the backlog state $q$ through Eq.~\eqref{eq:energy_used}.
Unless $k_q^\star=0$, the only contribution to the energy consumption that depends on $q$ is the energy due to processing.
Since $q E_P(k)$ increases with $q$, see Eq.~\eqref{eq:e_processing}, it is $k_{q,E}^\star(u) \le k_{1,E}^\star(u)$, i.e., given the amount of energy available, fitting more packets into a single slot is more expensive than processing a single packet.
The actual relation between $k_q^\star$ and $k_1^\star$ is not clear, because it highly depends on the energy the node allocates. If $k_{q,E}^\star(u) \ge k_{q,R}^\star$, then $k^\star_q = k_{q, R}^\star$ and the improvement stated by Theorem~\ref{th:retx} holds (whatever $k_1^\star$ is).
If instead $k_{q,E}^\star(u) < k_{q,R}^\star$, i.e., $k^\star_q = k_{q, E}^\star$, there is a simple relation between the average distortion obtained with and without the retransmission scheme and, according to the actual values of $k^\star_q$ and $k^\star_1$, it may be better to use the retransmission scheme or the single-transmission one\footnote{Following the rasoning of Appendix~\ref{apx:proof_retx}, if $k^\star_q \ge k^\star_1/q$, then the retransmission scheme leads to some improvement, otherwise nothing can be inferred, in general, because the performance depends on the particular shape of $\mathbb{E}_[\Delta]$ in the two cases.}.
Anyway, $k_{q,E}^\star(u)$ depends on the energy that the node decides to use in the considered time slot, and it is up to the MDP to manage it in such a way to obtain the lowest average distortion which, as shown in Theorem~\ref{th:retx}, is smaller when the retransmission scheme is adopted.
In practice, the retransmission mechanism coupled with an intelligent energy management scheme achieves a better QoS, i.e., lower distortion. 

As stated in Theorem~\ref{th:retx}, the retransmission scheme brings enhancements with respect to the single transmission one. However, this improvement is not proportional to the value of $r$, but rather depends on the particular shapes of the distortion and outage probability functions, and on how large $r$ is.
In particular, this is formalized as follows:~
\begin{thm} \label{th:retx_2}
 Consider an integer $r>2$. 
 The long-term average distortion achieved when using the retransmission scheme with $r$ may be lower than that obtained with $r-1$. 
 \begin{proof}
  See Appendix~\ref{apx:proof_retx_2}.
 \end{proof}
\end{thm}

In the proof, we showed that the probability of having worse performance with $r$ rather than $r-1$ increases as $r$ becomes larger.
Hence, $r$ should be kept reasonably small.
This is an intuitive result, since increasing $r$ implies reserving fewer transmitted bits per data block in case of retransmission, thereby introducing a larger compression and, consequently, distortion at the source. A deeper understanding of how the average distortion changes as $r$ increases is given in Appendix~\ref{apx:proof_retx_2}.

There are two additional factors that need to be taken into account in the design of the retransmission scheme.
\begin{itemize}
 \item \textit{Latency}: the information contained in a packet may lose significance as latency increases. In this case, sending a data block long after it has been originated may even be disadvantageous: receiving it brings no benefit to the final application, and, at the same time, reduces the quality of the other data blocks that are transmitted along with it, as it occupies part of the bits available for the transmission. 
 \item \textit{QoS}: the final application may dictate a minimum QoS threshold, i.e., a maximum distortion that can be tolerated on the received information. When data blocks are transmitted together, they are compressed more and have a larger distortion, which may violate the QoS constraints.
\end{itemize}
Thus, the choice of $r$ should be guided by the specific application constraints and by the values of $k_r^\star$ and $D(k^\star_r)$.

\section{Numerical evaluation} \label{sec:results}

In the following, we show how the average long-term cost $C^\star$ is affected by the system parameters and compare the performance obtained with the single-transmission scheme (i.e., $r=1$), and the retransmission scheme with $r=2$.
We also evaluate the average distortion obtained when an energy-unaware greedy scheme is adopted.
This \emph{greedy} policy is myopic and does not optimize the energy consumption according to the expected future availability; i.e., when in state $(x,b,q)$, the node uses all the energy it has in the battery, unless it is more than $u_q^\star$, i.e., the energy needed to achieve $k_{q,R}^\star$. Hence, $u= \min(b, u_q^\star)$.

We investigated the role of the system parameters by running RVIA for the chosen system configurations.
In all cases, the original packet size is $L_0=500$ bits and the nodes can decide among $m=30$ different compression ratios. 
The parameters of the distortion curve of Eq.~\eqref{eq:dist} have been derived from~\cite{Zordan-2017}; in particular, we set $a=0.35$ and $b=19.9$.
Notice that with these choices of $a,b$ and $m$, the granularity of the compression ratio (i.e., $1/m$) is such that $D(1) < D_\fl$, which guarantees that the distortion function is decreasing.

The considered  transmission power is 
$P_\ttx=25$ mW, the transmission frequency is $f_0=868.3$~MHz, the used bandwidth is $W = 125$~kHz, and the overall noise power spectral density $N_0=-167$~dBm/Hz.
The path loss exponent is $3.5$ and Rayleigh fading is modeled as an exponential random variable with unit mean.
The battery dynamics of Eq.~\eqref{eq:battery} assumes that the energy is quantized.
Consistently, all the terms of energy consumption, i.e.,  Eqs.~\eqref{eq:e_processing},~\eqref{eq:e_tx},~\eqref{eq:e_circuitry}, have been mapped into quanta (we ensured an appropriate granularity for this purpose).
In our numerical evaluation, the energy due to processing (see also~\cite{Zordan-2014}) is of the same order of magnitude of that needed for a transmission, whereas the circuitry contribution is smaller.
We remark that, according to Eq.~\eqref{eq:energy_used}, a packet cannot be sent if the available energy is below a certain threshold.
We also denote as $e_{\max}$ the maximum energy consumption demanded by the processing and transmission of a (single) packet, and introduce the normalized quantities $\overline{\mu} = \mu/e_{\max}$, and $\overline{B}=B/e_{\max}$. 

\begin{figure}[t]
  \setlength\fwidth{0.88\columnwidth}
  \setlength\fheight{0.7\columnwidth}
%
%
\definecolor{mycolor1}{rgb}{0.00000,0.44700,0.74100}%
\definecolor{mycolor2}{rgb}{0.85000,0.32500,0.09800}%
\definecolor{mycolor3}{rgb}{0.92900,0.69400,0.12500}%
\definecolor{violet}{rgb}{0.6,0,0.6}%
\definecolor{orange_D}{rgb}{1.0000,0.5,0}%
\definecolor{cyan}{rgb}{0,0.67,0.64}%
\begin{tikzpicture}
\pgfplotsset{every tick label/.append style={font=\scriptsize}}
\tikzstyle{dotted}= [dash pattern=on \pgflinewidth off 0.5mm] 
\tikzstyle{dashed}= [dash pattern=on 7.5*0.8*0.8pt off 7.5*0.4*0.8pt]
\tikzstyle{dashdotted} = [dash pattern=on 7.5*0.8*0.6pt off 7.5*0.8*0.3pt on \the\pgflinewidth off 7.5*0.8*0.3pt]
\tikzstyle{dotted2} = [dash pattern=on 7.5*0.8*0.3pt off 7.5*0.8*0.2pt]

\begin{axis}[%
width=0.955\fwidth,
height=0.485\fheight,
at={(0\fwidth,0.515\fheight)},
scale only axis,
xmin=25,
xmax=500,
xtick={100,200,300,400,500},
xticklabels={,,,,},
xlabel style={font=\footnotesize\color{white!15!black}},
ymin=0.1,
ymax=1.05,
ylabel style={font=\footnotesize\color{white!15!black}},
ylabel={$C^\star$},
axis background/.style={fill=white},
xmajorgrids,
ymajorgrids,
legend style={font=\footnotesize, at={(0.98,0.043)}, anchor=south east, legend cell align=left, align=left, draw=white!15!black}
]
\addplot [color=orange_D, line width=1pt,  mark=triangle*, mark size=1.3, mark repeat=2
]
  table[row sep=crcr]{%
25	0.253880056917084\\
50	0.255441389887147\\
75	0.259620077520261\\
100	0.292529609826858\\
125	0.354910900498214\\
150	0.446048204528333\\
175	0.565265251509292\\
200	0.711346971149565\\
225	0.827661427459551\\
250	0.914345750081329\\
275	0.971654010209204\\
300	1\\
325	1\\
350	1\\
375	1\\
400	1\\
425	1\\
450	1\\
475	0.999999999999999\\
500	0.999999999999999\\
};
\addlegendentry{single-tx}

\addplot [color=cyan, line width=1pt,  mark=*, mark size=1.3, mark repeat=2, mark phase=2]  
table[row sep=crcr]{%
25	1\\
50	1\\
75	1\\
100	1\\
125	1\\
150	1\\
175	1\\
200	1\\
225	1\\
250	1\\
275	1\\
300	1\\
325	1\\
350	1\\
375	1\\
400	1\\
425	1\\
450	1\\
475	1\\
500	1\\
};
\addlegendentry{retx r=2}

\addplot [color=violet, line width=1,  mark=square*, mark size=1.3, mark repeat=2, mark phase=2]
  table[row sep=crcr]{%
25	0.258076086840555\\
50	0.259628639185031\\
75	0.263783826711343\\
100	0.296508282364239\\
125	0.358538753083993\\
150	0.449163519071938\\
175	0.567710113130203\\
200	0.712970298280525\\
225	0.828630625259519\\
250	0.914827452519474\\
275	0.971813422408911\\
300	1\\
325	1\\
350	1\\
375	1\\
400	1\\
425	1\\
450	1\\
475	1\\
500	1\\
};
\addlegendentry{greedy}

\end{axis}

\begin{axis}[%
width=0.955\fwidth,
height=0.485\fheight,
at={(0\fwidth,0\fheight)},
scale only axis,
xmin=25,
xmax=500,
xtick={100,200,300,400,500},
xticklabels={100,200,300,400,500},
xlabel style={font=\footnotesize\color{white!15!black}},
xlabel={$d$ [m]},
ymin=0.1,
ymax=0.55,
ytick={0.1,0.2,0.3,0.4,0.5},
yticklabels={0.1,0.2,0.3,0.4,0.5},
ylabel style={font=\footnotesize\color{white!15!black}},
ylabel={$C^\star$},
axis background/.style={fill=white},
xmajorgrids,
ymajorgrids,
legend style={font=\footnotesize at={(0.97,0.03)}, anchor=south east, legend cell align=left, align=left, draw=white!15!black}
]
\addplot [color=orange_D, line width=1pt,  mark=triangle*, mark size=1.3, mark repeat=2
]
  table[row sep=crcr]{%
25	0.250044806736902\\
50	0.251614165363712\\
75	0.255814332535936\\
100	0.261473839124158\\
125	0.269899697255378\\
150	0.281002846828115\\
175	0.294772052274433\\
200	0.310666967119474\\
225	0.327284247891832\\
250	0.344318086555363\\
275	0.361436777815066\\
300	0.378379185309514\\
325	0.395056777867221\\
350	0.411841152587242\\
375	0.428540487199022\\
400	0.445082808145469\\
425	0.461466020826164\\
450	0.477662190107212\\
475	0.49377480751933\\
500	0.509680295428368\\
};

\addplot [color=cyan, line width=1pt,  mark=*, mark size=1.3, mark repeat=2, mark phase=2]
  table[row sep=crcr]{%
25	0.196076223961439\\
50	0.196493736336253\\
75	0.197663928594241\\
100	0.200014533922853\\
125	0.204131683680844\\
150	0.210307304276266\\
175	0.21894993500336\\
200	0.230301059011903\\
225	0.242982706761126\\
250	0.256892164440783\\
275	0.271811502732548\\
300	0.287301506206019\\
325	0.303153094348552\\
350	0.31941005442526\\
375	0.336026121329815\\
400	0.352769504474414\\
425	0.369407121505962\\
450	0.386065558173767\\
475	0.402768757853503\\
500	0.419799257683124\\
};

\addplot [color=violet, line width=1,  mark=square*, mark size=1.3, mark repeat=2, mark phase=2]
  table[row sep=crcr]{%
25	0.268076086840555\\
50	0.269628639185031\\
75	0.273783826711343\\
100	0.279336567861388\\
125	0.287444488252377\\
150	0.297996495506656\\
175	0.310960858592059\\
200	0.325805124470814\\
225	0.34123644666662\\
250	0.357228207224187\\
275	0.373503373628691\\
300	0.38985002372592\\
325	0.40618170757277\\
350	0.4228194390276\\
375	0.439432875382099\\
400	0.455923308006418\\
425	0.472263306722704\\
450	0.488422029597565\\
475	0.504498663509046\\
500	0.520372437250718\\
};

\end{axis}

\end{tikzpicture}%
  \caption{Average long-term distortion as a function of the distance $d$ for $\bar{B}=0.6$ (top) and $\bar{B}=0.8$ (bottom) when $\bar{\mu}=1$.}
  \label{fig:dist}
\end{figure}
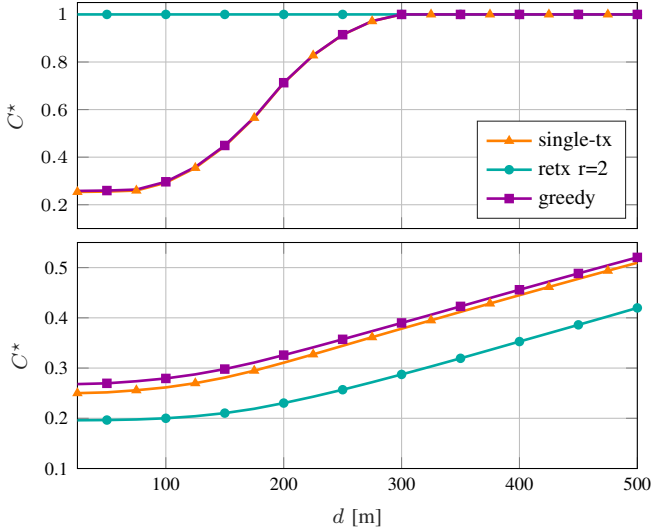

In the energy harvesting process, the probability that the source goes from the bad to the good state is 3 times greater than that of the opposite transition.
In the bad state ($x=0$), $e=0$ quanta of energy arrive with probability 1, otherwise the energy inflow follows a Gaussian distribution with mean $\mu$ and variance 10.

Fig.~\ref{fig:dist} shows the average long-term cost as a function of the distance from the receiver when $\bar{\mu}=1$ and for two different values of the battery size, namely $\bar{B}=0.6$ and $\bar{B}=0.8$.
Clearly, when the node is farther from its receiver, the path loss component increases, thereby leading to a larger outage probability.
To guarantee that its measurements do not get lost in the transmission, the node will use a stronger coding rate, thus the compression ratio needs to be smaller and a larger distortion is introduced at the source. 
In both cases, the battery size is quite small and such that the node's readings generally have to be compressed because the energy stored is not enough to send them as they are.
This  clearly impinges on the distortion that can be achieved on average. What is interesting to note is the behavior of the retransmission policy when $\bar{B}=0.6$ (Fig.~\ref{fig:dist}, top).
As discussed in Section~\ref{ssec:retx_eff}, the energy required to send two or more packets together is larger than that needed for a single one, since it requires to spend more energy for processing. 
In this case, the battery size is very small, and the node cannot store enough energy to transmit two packets together.
In the end, the node will decide to almost always discard the packets, or compress them a lot to be more robust against failures.
This result corroborates the impact of the energy availability on the performance.
Also, because of the small battery size, the optimal single-transmission policy performs similarly to the greedy one.
When instead $B$ is large enough, the retransmission mechanism leads to improvements with respect to the single-transmission one, as depicted in Fig.~\ref{fig:dist}, bottom.

\begin{figure}[t]
  \setlength\fwidth{.88\columnwidth} 
  \setlength\fheight{0.85\columnwidth}  
  \input{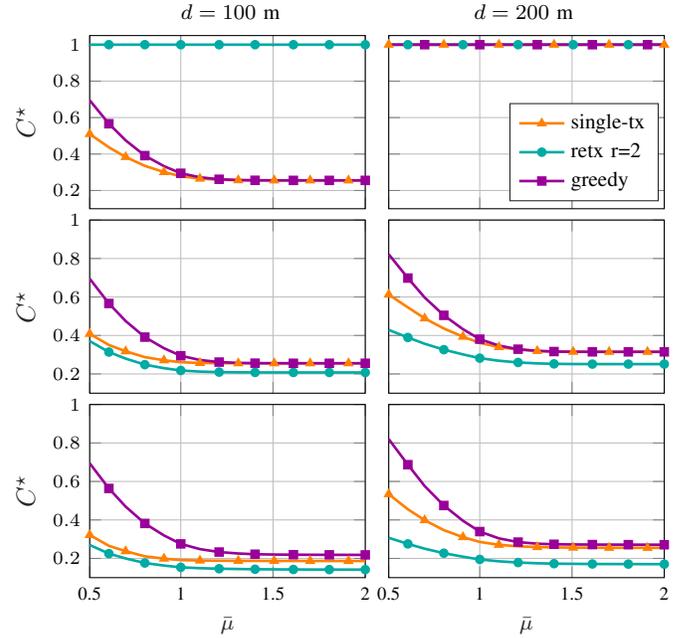}
  \caption{Average long-term distortion as a function of the average energy income during the ``good'' state $\bar{\mu}$ for $\bar{B}=0.6$ (top), $\bar{B}=0.8$ and $\bar{B}=2$ (bottom) and for $d=100$~m (left) and $d=200$~m (right).}
  \label{fig:mu}
\end{figure}

Fig.~\ref{fig:mu} shows the performance as a function of the average energy income during the good source state, $\mu$.
The distance of the node is fixed to $100$~m for the figures on the left, and $200$~m for those on the right.
The normalized battery size is $\bar{B} = 0.6, 0.8,$ and $2$ for the figures on the top, middle, and bottom, respectively.
Intuitively, the more the energy that can be harvested (large $\mu$), the lower the average distortion, because the node can choose the optimal point in the rate distortion tradeoff (see Sec.~\ref{ssec:RDP}) more often.
However, after a certain value of $\mu$, the distortion curve tends to remain constant because (i) the battery size is too small and part of the incoming energy needs to be discarded, and (ii) the optimal $k^\star$ is already achievable and the exceess energy is not useful.
As already seen in Fig.~\ref{fig:dist}, if the battery size is too small compared to the minimum energy required to send $q=1,\dots,r$ packets, the node will never transmit them, and this effect sharpens with the distance.
Fig.~\ref{fig:dist} also shows that, when the node is closer to the receiver and there is enough energy available (second and third plots on the left), the single-transmission optimal policy behaves similarly to the retransmission one, as the outage probability is low; otherwise, its performance becomes closer to that of the greedy policy and the improvement obtained with the retransmission scheme becomes more significant (see plots on the right).

\begin{figure}[t]
  \setlength\fwidth{.9\columnwidth} 
  \setlength\fheight{0.5\columnwidth}  
%
%
\definecolor{mycolor1}{rgb}{0.00000,0.44700,0.74100}%
\definecolor{mycolor2}{rgb}{0.85000,0.32500,0.09800}%
\definecolor{mycolor3}{rgb}{0.92900,0.69400,0.12500}%
\definecolor{mycolor4}{rgb}{0.49400,0.18400,0.55600}%
\definecolor{mycolor5}{rgb}{0.46600,0.67400,0.18800}%
\definecolor{mycolor6}{rgb}{0.30100,0.74500,0.93300}%

\definecolor{violet}{rgb}{0.6,0,0.6}%
\definecolor{orange_D}{rgb}{1.0000,0.5,0}%
\definecolor{cyan}{rgb}{0,0.67,0.64}%
\definecolor{red}{rgb}{0.9,0,0}%
\definecolor{green}{rgb}{0,0.8,0}%
\begin{tikzpicture}
\pgfplotsset{every tick label/.append style={font=\scriptsize}}
\tikzstyle{dotted}= [dash pattern=on \pgflinewidth off 0.5mm] 
\tikzstyle{dashed}= [dash pattern=on 7.5*0.8*0.8pt off 7.5*0.4*0.8pt]
\tikzstyle{dashdotted} = [dash pattern=on 7.5*0.8*0.6pt off 7.5*0.8*0.3pt on \the\pgflinewidth off 7.5*0.8*0.3pt]
\tikzstyle{dotted2} = [dash pattern=on 7.5*0.8*0.3pt off 7.5*0.8*0.2pt]

\begin{axis}[%
width=0.955\fwidth,
height=\fheight,
at={(0\fwidth,0\fheight)},
scale only axis,
xmin=25,
xmax=500,
xlabel style={font=\footnotesize\color{white!15!black}},
xlabel={$d$ [m]},
ymin=0.1,
ymax=1.05,
ylabel style={font=\footnotesize\color{white!15!black}},
ylabel={$C^\star$},
axis background/.style={fill=white},
xmajorgrids,
ymajorgrids,
legend style={font=\scriptsize, at={(0.27,0.91)}, anchor=north east, legend cell align=left, align=left, draw=white!15!black}
]
\addplot [color=orange_D, line width=0.8pt,  mark=triangle*, mark size=1.5, mark repeat = 2, mark phase = 1]
  table[row sep=crcr]{%
25	0.250044806736902\\
50	0.250840753805505\\
75	0.25395693554873\\
100	0.258392254579153\\
125	0.265836912009389\\
150	0.278342339678013\\
175	0.292970044461604\\
200	0.309472683412416\\
225	0.327238281810741\\
250	0.345311486234597\\
275	0.3614287435398\\
300	0.378139237779261\\
325	0.395066139710929\\
350	0.411950319282986\\
375	0.428699449023128\\
400	0.445350617139909\\
425	0.46163591083816\\
450	0.477777744443162\\
475	0.493866382686461\\
500	0.509680295428368\\
};
\addlegendentry{r = 1}

\addplot [color=cyan, line width=0.8pt,  mark=*, mark size=1.5, mark repeat = 2, mark phase = 2]
  table[row sep=crcr]{%
25	0.196076223961439\\
50	0.196285855863157\\
75	0.197119129184162\\
100	0.198344920373725\\
125	0.20049351358872\\
150	0.207829250604502\\
175	0.21687160465311\\
200	0.227997232161273\\
225	0.241558074009196\\
250	0.257249133631433\\
275	0.271237489350616\\
300	0.286418893051397\\
325	0.302593923620098\\
350	0.319008091376551\\
375	0.336507074344098\\
400	0.352522289734158\\
425	0.369499227574168\\
450	0.386310839343096\\
475	0.402196176534288\\
500	0.419799257683124\\
};
\addlegendentry{r = 2}

\addplot [color=violet, line width=0.8pt,  mark=x, mark size=2, mark repeat = 2, mark phase = 1]
  table[row sep=crcr]{%
25	0.163539982263416\\
50	0.163736020055795\\
75	0.164511963119541\\
100	0.165632056376904\\
125	0.167548656886543\\
150	0.17470437114504\\
175	0.18332024164838\\
200	0.193819921241826\\
225	0.206652775832553\\
250	0.221660514756302\\
275	0.235035956040672\\
300	0.249717241893646\\
325	0.265901014985286\\
350	0.282297797042207\\
375	0.300784589643603\\
400	0.317217492842937\\
425	0.334593191765463\\
450	0.351562949737887\\
475	0.367274129023381\\
500	0.384743182252112\\
};
\addlegendentry{r = 3}

\addplot [color=red, line width=0.8pt,  mark=|, mark size=1.8, mark repeat = 2, mark phase = 2]
  table[row sep=crcr]{%
25	0.149333488598446\\
50	0.149510685525785\\
75	0.150211124680722\\
100	0.15122042611507\\
125	0.15294287720552\\
150	0.160083416708946\\
175	0.168501377119194\\
200	0.178656907124915\\
225	0.190989012687543\\
250	0.205402002654\\
275	0.217824244682\\
300	0.23175730398373\\
325	0.247244338224705\\
350	0.263581705096545\\
375	0.281794385465064\\
400	0.29822472060665\\
425	0.315748187498582\\
450	0.333266110872955\\
475	0.34914379462915\\
500	0.36778680207682\\
};
\addlegendentry{r = 4}

\addplot [color=green, line width=0.8pt,  mark=diamond*, mark size=1.5, mark repeat = 2, mark phase = 1]
  table[row sep=crcr]{%
25	0.142903602094817\\
50	0.143074721251602\\
75	0.143751146717664\\
100	0.144725935909146\\
125	0.146389700426334\\
150	0.153555749822204\\
175	0.161940258717031\\
200	0.171996607089983\\
225	0.184151227100732\\
250	0.198315338120088\\
275	0.210297376280304\\
300	0.223722684118947\\
325	0.238747420202238\\
350	0.254472097406917\\
375	0.272364539824326\\
400	0.288793405682157\\
425	0.307558431650869\\
450	0.32597949380359\\
475	0.34370796002924\\
500	0.362974364204358\\
};
\addlegendentry{r = 5}

\addplot [color=mycolor6, line width=0.8pt,  mark=square*, mark size=1.2, mark repeat = 2, mark phase = 1]
  table[row sep=crcr]{%
25	0.999999999999991\\
50	1\\
75	1\\
100	1\\
125	1\\
150	0.999999999999997\\
175	0.999999999999995\\
200	0.999999999999995\\
225	0.999999999999994\\
250	0.999999999999995\\
275	0.999999999999995\\
300	0.999999999999994\\
325	0.999999999999996\\
350	0.999999999999994\\
375	0.999999999999992\\
400	0.999999999999994\\
425	0.999999999999993\\
450	0.999999999999993\\
475	0.999999999999995\\
500	0.999999999999996\\
};
\addlegendentry{r = 6}

\end{axis}
\end{tikzpicture}%
  \caption{Average long-term distortion as a function of the distance $d$ for different values of $r$ when $\bar{B}=1$ and $\bar{\mu}=1$.}  
  \label{fig:r}
\end{figure}
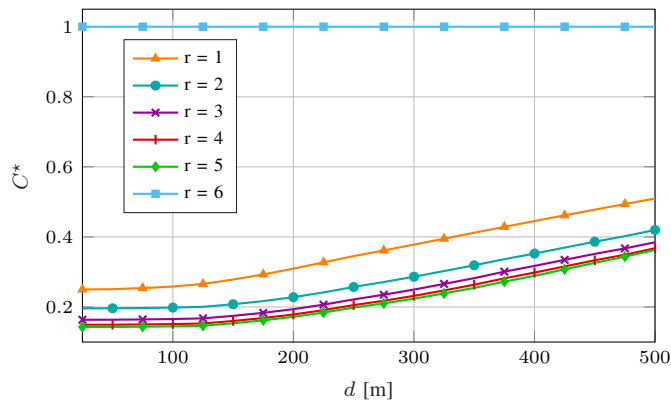

\begin{figure}[t]
  \setlength\fwidth{.9\columnwidth} 
  \setlength\fheight{0.5\columnwidth}  
%
%
\definecolor{mycolor1}{rgb}{0.00000,0.44700,0.74100}%
\definecolor{mycolor2}{rgb}{0.85000,0.32500,0.09800}%
\definecolor{mycolor3}{rgb}{0.92900,0.69400,0.12500}%
\definecolor{mycolor4}{rgb}{0.49400,0.18400,0.55600}%
\definecolor{mycolor5}{rgb}{0.46600,0.67400,0.18800}%
\definecolor{mycolor6}{rgb}{0.30100,0.74500,0.93300}%

\definecolor{violet}{rgb}{0.6,0,0.6}%
\definecolor{orange_D}{rgb}{1.0000,0.5,0}%
\definecolor{cyan}{rgb}{0,0.67,0.64}%
\definecolor{red}{rgb}{0.9,0,0}%
\definecolor{green}{rgb}{0,0.8,0}%
\begin{tikzpicture}
\pgfplotsset{every tick label/.append style={font=\scriptsize}}
\tikzstyle{dotted}= [dash pattern=on \pgflinewidth off 0.5mm] 
\tikzstyle{dashed}= [dash pattern=on 7.5*0.8*0.8pt off 7.5*0.4*0.8pt]
\tikzstyle{dashdotted} = [dash pattern=on 7.5*0.8*0.6pt off 7.5*0.8*0.3pt on \the\pgflinewidth off 7.5*0.8*0.3pt]
\tikzstyle{dotted2} = [dash pattern=on 7.5*0.8*0.3pt off 7.5*0.8*0.2pt]

\begin{axis}[%
width=0.955\fwidth,
height=\fheight,
at={(0\fwidth,0\fheight)},
scale only axis,
xmin=1,
xmax=11,
xtick={1,3,5,7,9,11},
xticklabels={0.5,0.7,0.9,1.1,1.3,1.5},
xlabel style={font=\footnotesize\color{white!15!black}},
xlabel={$\bar{\mu}$},
ymin=0,
ymax=1,
ylabel style={font=\footnotesize\color{white!15!black}},
ylabel={discarded probability},
axis background/.style={fill=white},
xmajorgrids,
ymajorgrids,
legend style={font=\scriptsize, at={(0.97,0.95)}, anchor = north east, legend cell align=left, align=left, draw=white!15!black}
]

\addplot [color = orange_D, line width=0.8pt,  mark=triangle*, mark size=1.5, mark repeat = 2, mark phase = 1]
  table[row sep=crcr]{%
1	0.455387469187041\\
2	0.375311667059363\\
3	0.324576791373048\\
4	0.277967083111073\\
5	0.249183318712655\\
6	0.23027397246817\\
7	0.220583391782766\\
8	0.216196287943793\\
9	0.214261246637988\\
10	0.213074751471137\\
11	0.212663721348145\\
};
\addlegendentry{r = 1}

\addplot [color = cyan, line width=0.8pt,  mark=*, mark size=1.5, mark repeat = 2, mark phase = 2]
  table[row sep=crcr]{%
1	0.220889763490828\\
2	0.204294068274602\\
3	0.191015431776272\\
4	0.178011079193772\\
5	0.167100048816021\\
6	0.157444885961542\\
7	0.14960063999461\\
8	0.143411923531056\\
9	0.139017906120624\\
10	0.135309351810433\\
11	0.133897913526166\\
};
\addlegendentry{r = 2}

\addplot [color = mycolor4, line width=0.8pt,  mark=x, mark size=2, mark repeat = 2, mark phase = 1]
  table[row sep=crcr]{%
1	0.181656369739554\\
2	0.170101539524042\\
3	0.159824046380832\\
4	0.14985340785156\\
5	0.141092712320673\\
6	0.133351947906666\\
7	0.127061185470565\\
8	0.122286340233503\\
9	0.119017502564461\\
10	0.116315521238068\\
11	0.115458988747985\\
};
\addlegendentry{r = 3}

\addplot [color = red, line width=0.8pt,  mark=|, mark size=1.8, mark repeat = 2, mark phase = 2]
  table[row sep=crcr]{%
1	0.15129513559141\\
2	0.138278647337192\\
3	0.128391071381958\\
4	0.119253006360771\\
5	0.112436307980205\\
6	0.107502795302853\\
7	0.104427980915169\\
8	0.10268188067138\\
9	0.101857270801896\\
10	0.101498867626301\\
11	0.101481761138051\\
};
\addlegendentry{r = 4}

\addplot [color = green, line width=0.8pt,  mark=diamond*, mark size=1.5, mark repeat = 2, mark phase = 1]
  table[row sep=crcr]{%
1	0.12762394920373\\
2	0.114360034879421\\
3	0.104284593702488\\
4	0.0949200603973359\\
5	0.08788415855539\\
6	0.0826441997929031\\
7	0.0792601519393512\\
8	0.0772143074564252\\
9	0.0761471399120271\\
10	0.0755529490492554\\
11	0.0754654449559806\\
};
\addlegendentry{r = 5}

\addplot [color = violet, line width=0.8pt,  mark=square*, mark size=1.2, mark repeat = 2, mark phase = 2]
  table[row sep=crcr]{%
1	0.853710913090851\\
2	0.699047062406781\\
3	0.574232016725612\\
4	0.460986372125632\\
5	0.374823379654074\\
6	0.315653260033823\\
7	0.279578165438913\\
8	0.26121620897413\\
9	0.253752449724753\\
10	0.250665844984439\\
11	0.250163893844067\\
};
\addlegendentry{greedy}

\end{axis}
\end{tikzpicture}%
  \caption{Packet loss probability as a function of $\bar{\mu}$ for different values of $r$ when $\bar{B}=1$ and $d=200$m.}  
  \label{fig:discarded}
\end{figure}
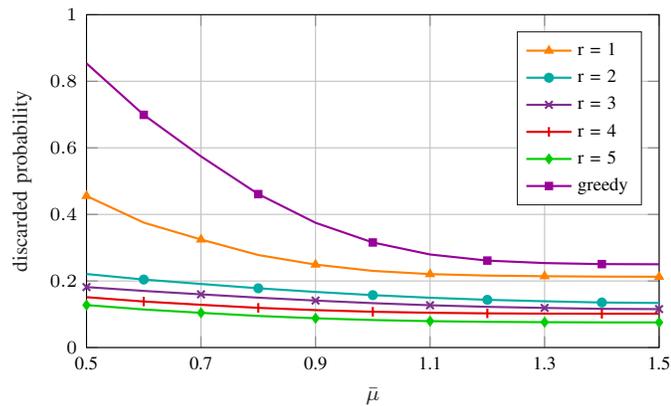

The effect of the maximum number of retransmissions on the distortion as a function of distance is shown in Fig.~\ref{fig:r} for $\bar{B}=1$ and $\bar{\mu}=1$.
As $r$ increases, the gain obtained becomes smaller: if many packets are sent together in a single slot, the distortion introduced by the lossy compression is large, and this reduces the benefit of retransmitting.
If $r$ becomes too large, the number of channel uses available for each packet is so small that the compression ratio that is required is too large and this affects the dynamics of the MDP and thus of the system performance.

Finally, Fig.~\ref{fig:discarded} shows the packet loss probability as a function of $\mu$ for different values of $r$ and $\bar{B}=1$, where this probability accounts for both the packets lost in the communication and those discarded at the source.
Again, it can be observed how the benefits of the retransmission scheme fade as $r$ increases. 
The gain obtained over the greedy policy by the Value Iteration algorithm is more significant when energy is scarce (small $\mu$), for both the single-transmission scheme ($r=1$) and the retransmission mechanism ($r>1$).

%
%


\section{A learning approach} \label{sec:learning}

The policies found with the RVIA are optimal and have a threshold structure, cf. \cite{Pielli-2017}, which implies limited storage requirements and low implementation complexity.
However, this approach requires the knowledge of the statistics of the MDP, which means that the dynamics of the EH process and the average channel gain need to be known, but this information may not be available with enough accuracy.
A natural way to determine the optimal policy when the knowledge about the environment is uncertain or limited is to interact with it and learn based on the feedback received.
This approach is commonly referred to as \emph{reinforcement learning}~\cite{Sutton-98} and is usually applied to scenarios where the environment is modeled through an MDP, which is the case in this paper.
In particular, in this paper we exploit the R-learning algorithm, which is the counterpart of the well-known Q-learning algorithm for the average case rather than for discounted optimization problems.

\subsection{R-learning} \label{ssec:R-learning}

R-learning is an RL algorithm introduced by Schwartz in 1993~\cite{schwartz-1993} for average-reward maximization problems.
Notice that it can be promptly adapted to our model by using the negatives of the costs. 
The basic gist of R-learning is to learn the value of each admissible action for all possible states through exploration.
As discussed in Section~\ref{ssec:RVI}, in the long run the average cost, see Eq.~\eqref{eq:J}, is the same regardless of the initial state.
However, there is a transient, and it is this transient that defines the value of a state-action pair:~
\begin{equation}
 q^\pi(s,u) = \sum\limits_{k=1}^\infty \mathbb{E}\left[R_{n+k} - J^\pi \middle | s_n=s,u_n=u \right],
\end{equation}
where $R_n = -c(s_n, u_n, s_{n+1})$ is the reward obtained in slot $n$; notice that the next state $s_{n+1}$ depends on action $u_n$ and on the environment. The function $q^\pi$ represents the value relative to the average reward under the current policy.
The essential approach of RL algorithms is \emph{exploration vs exploitation}: in  a state, an agent can choose the action that leads to the highest reward based on current information ({exploitation}), or keep trying new actions, hoping that they bring even higher rewards ({exploration}).

Similarly to Q-learning, R-learning maintains (i) a behavior policy that simulates experience and dictates the action to choose, (ii) an estimation policy $\pi$ which is the one involved in the policy iteration process and is the policy that is being learned, (iii) an action-value function $Q^L$ that approximates $q^\pi$, and (iv) an estimated average reward $\rho$ that approximates $J^\pi$.
Notice that, in order to learn the estimation policy $\pi$, the device has to use an additional policy (the behavior one), that dictates what to do during the learning process. 
We chose to use an $\varepsilon$-greedy policy as behavior policy: when in state $s$, the node chooses a random action $u\in\mathcal{U}_s$ with probability $\varepsilon$, otherwise it selects the action with the highest $Q^L$ value, i.e., the current best action. 

\begin{algorithm}[!t]
\caption{R-learning algorithm}\label{alg:R-learning}
\begin{algorithmic}[1]
\State Initialize $\rho$ and $Q^L(s,u) \:\forall s,u$
\While {true}
\State $s \gets$ current state
\State Choose action $u\in\mathcal{U}_s$ using $\varepsilon$-greedy policy
\State Observe next state $s'$ and reward $R$
\State $\delta \gets R - \rho + \max_w Q^L(s',w) - Q^L(s,u)$
\State $Q^L(s,u) \gets Q^L(s,u) + \alpha \delta$
\If {$Q^L(s,u) = \max_w Q^L(s,w)$}
\State $\rho \gets \rho + \beta \delta$
\EndIf
\EndWhile
\end{algorithmic}
\end{algorithm}

The complete algorithm is given in Algorithm~\ref{alg:R-learning}~\cite{Sutton-98}.
The scalars $\alpha$ and $\beta$ are step-size hyperparameters.
During the learning process, the node continuously updates the $Q^L$ values of all state-action pairs and, at each iteration, chooses either the action corresponding to the highest $Q^L$ value or a random one, according to the $\varepsilon$-greedy policy.
After some iterations, the result of $\max_{u\in\mathcal{U}_s} Q^L(s,u)$ will converge and the node learns that the best action to take in state $s$ is:~
\begin{equation}
 u^L(s) = \max_{u\in\mathcal{U}_s} Q^L(s,u). \label{eq:u_L}
\end{equation}
Hence the policy learned by the device coincides with the \mbox{$0$-greedy} one, and in each state $s\in \mathcal{S}$ the action to take is deterministic and given by~\eqref{eq:u_L}.

\begin{figure}[t]
  \setlength\fwidth{.9\columnwidth} 
  \setlength\fheight{0.5\columnwidth}  
  \input{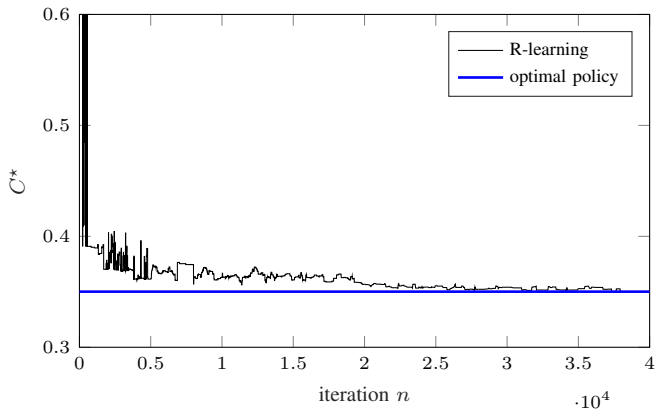}
  \caption{Average long-term distortion obtained with Algorithm~\ref{alg:R-learning} as a function of the number of learning iterations, and comparison with the corresponding optimal cost. Here, $\rho=5$ and $\beta = 0.01$.}  
  \label{fig:learning}
\end{figure}

The convergence of the learning algorithm is demonstrated in Fig.~\ref{fig:learning}, which shows the average long-term cost of Eq.~\eqref{eq:final_cost} as the node is learning the policy.
The simulation parameters in this case are $r=2$, $\bar{B}=0.8, \bar{\mu} = 0.7$, and $d=200$~m.
Fig.~\ref{fig:learning} shows the average long-term cost obtained with the policy learned by the learning algorithm vs the number of iterations of the learning phase.
The blue line represents the distortion obtained with the optimal policy of Section~\ref{sec:solution}.
The hyperparameters of the learning algorithm were optimized by performing an exhaustive search and selecting the combinations that performed best in the training phase.
Both the exploration rate $\varepsilon$ and the learning rate $\alpha$ were chosen to be decreasing during the learning process.
It is a reasonable and common practice that, as the learning proceeds and the $Q^L$-values converge, $\varepsilon$ decreases. This happens because initially the device has no knowledge about the environment and makes random moves to maximally explore the state space, so that all possible states can eventually be visited, and the long-term $Q^L$-value of every state-action pair can be determined. Then, it converges to a small exploration rate and exploits the accumulated knowledge~\cite{Sutton-98}.





\section{Conclusions} \label{sec:conclu}

In this paper, we proposed and analyzed a joint retransmission, compression and channel-coding scheme for energy-constrained sensors that access the channel in a TDMA fashion.
The goal is to enable the sensors to be energetically self-sufficient and, at the same time, to minimize the long-term average distortion of the data they report to the receiver.
The proposed scheme foresees that the data blocks that were not successfully reported are retransmitted along with the newly generated blocks, where the compression is performed separately for all blocks, while the channel-coding procedure is performed for all blocks jointly. 
The problem has been formulated by means of an MDP, and the optimal policy was derived with a variant of the Value Iteration algorithm.
We also proposed the use of a learning approach to determine the policy during the operation of the system. 

Both the analytical investigation and the numeral evaluation showed that the retransmission scheme ensures an improved average quality of the received information with respect to the simpler single-transmission scheme, unless the energy available to the node is very scarce (because either the battery size is too small or the energy inflow is insufficient).
In this case, the retransmission scheme demands too much energy and the node cannot afford it.
When the energy is scarce, a more flexible scheme that decides whether or not to retransmit a packet may improve the performance. For instance, if a packet is lost and the battery charge is low, then the node should opt for transmitting only the new readings and give up on those that were lost, so as to preserve some energy.
The gain introduced by the retransmission scheme becomes more significant as the probability of outage increases. 
As the maximum number of transmission attempts that can be dedicated to a packet increases, the performance improvement tends to fade, while also affecting latency. 
We also validated the effectiveness of machine learning algorithms, which can be extremely powerful when the information available to the node is imprecise or incomplete, and achieve a performance very close to that of the optimal policies.

 In this work, it was assumed that a node can retransmit its lost packets only in the time slots dedicated to the sensor device.
An interesting alternative consists in considering some shared time slots, where the devices can send their lost data in a random fashion, i.e., contending for the channel. A study of this scheme is part of our future work.

In our further work, we also intend to adopt a realistic model for the battery consumption, which in practice is not linear, unlike commonly assumed, but depends on the current battery level.
Another extension is the investigation 
of how the actual performance changes when the knowledge about the average channel state is imperfect.
Finally, it would be interesting to generalize the model to other compression techniques, possibly with different energy consumption characteristics.


\section*{Acknowledgments}
The work of C. Pielli and M. Zorzi was partly supported by Intel's Corporate Research Council, under the project ``EC-CENTRIC: an energy- and context-centric optimization framework for IoT nodes''. The work of P. Popovski and C. Stefanovic was in part supported by the European Research Council (ERC Consolidator Grant Nr. 648382 WILLOW) within the Horizon 2020 Program.

\appendices

\section{Proof of Theorem~\ref{th:retx}}\label{apx:proof_retx}

We compare the case in which a packet that is not received cannot be retransmitted ($r=1)$ with that in which a data block that was lost can be combined together with the next data blocks and retransmitted for at most $r>1$ times.

The proof leverages on the fact that the distortion function is convex and decreasing.
We will denote $D_\fl$ as $D(0)$.
By assumption, we neglect the energy constraints, i.e., $k_r^\star = k_{r,R}^\star$ and $k_1^\star = k_{1,R}^\star$, see Eq.~\eqref{eq:k_star}.

\textbf{Step 1.} First of all, we prove that $k_{r}^\star \ge k_{1}^\star/r$. 

\begin{figure}
 \centering
 \setlength\fwidth{.8\columnwidth}
 \setlength\fheight{0.47\columnwidth}
 \input{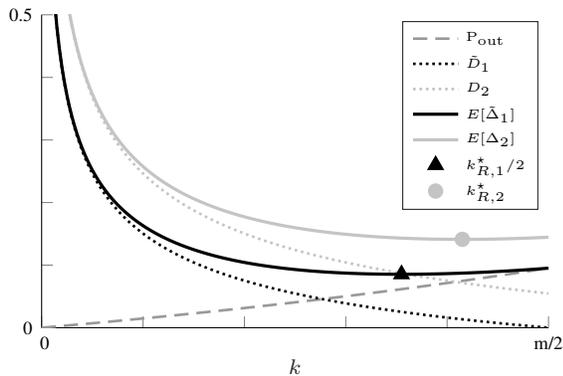}
 \caption{Example of locations of $k_{R,2}^\star$ and $k_{1,R}^\star/2$. The tilde notation indicates that the function is scaled from $[0, m]$ to $[0, m/2]$.}
 \label{fig:proof_dist}
\end{figure}

\noindent When $r>1$ and $r-1$ consecutive transmissions failed, the lost $r-1$ data blocks are retransmitted in the successive time slot along with the most recently generated data block.
In this case, 
the device selects $k_r^\star \in [0,m/r]$ as the optimal point in the rate-distortion tradeoff.
According to our assumption, $k_r^\star =k_{r,R}^\star$, which is the point of minimum of function $\mathbb{E}[\Delta_r(k)]$, see Eq.~\eqref{eq:delta}.
The outage probability depends on the coding rate $R=rL(k)/S$ and scales with $r$, 
i.e., it maintains the same shape regardless of $r$ because $k$ cannot exceed $m/r$. The outage probability is represented with a dashed line in Fig.~\ref{fig:proof_dist} for $k\in[0,m/r]$ with $r=2$.
If the distortion introduced at the source also scaled with $r$, it would be $k_{r}^\star = k_1^\star/r$; instead, it only depends on $L(k)$ and is truncated to $D(m/r)$. Since $D(k)$ is a decreasing function of the compression ratio, the effective distortion in the interval $[0, m/r]$ (light dotted line in Fig.~\ref{fig:proof_dist}) is larger than the one that would be obtained by scaling the original distortion in $[0, m]$ to the new range (dark dotted line in Fig.~\ref{fig:proof_dist}). 
This means that the point of minimum of $\mathbb{E}[\Delta_r(k)]$ is shifted right with respect to that obtained when simply scaling both the outage probability and the distortion functions from $[0,m]$ to $[0,m/r]$. 
Then:~
\begin{equation} \label{eq:k_r}
 k_{r}^\star \ge k_{1}^\star/r. 
\end{equation}
An example of this can be seen in Fig.~\ref{fig:proof_dist}: the light and dark flat curves show  $\mathbb{E}[\Delta_2(k)]$ and the scaled version of $\mathbb{E}[\Delta_1(k)]$, respectively, and the markers identify their minima. 

\textbf{Step 2.} We show that $(r-1)\,D(0) + D(k_1^\star) \ge r\,D(k_r^\star)$.

\noindent As a first step, focus on $r=2$. The convexity of $D(k)$ implies that:~
\begin{align}
 D(x_1) &+  D(x_2) \ge \\
 & D\left(\lambda x_1 + (1-\lambda) x_2\right) + D\left((1-\lambda) x_1 + \lambda x_2\right), \label{eq:convexity} 
 \end{align}
%
for any $\lambda\in[0,1]$. In particular, given that $D(k)$ is a decreasing function of $k$, for $x_1 < x_2$ and $\lambda \ge 1/2$, we also have:~
\begin{align}
 D(x_1) + D(x_2) & \ge \\
 & 2 D\left(\lambda x_1 + (1-\lambda) x_2\right) \triangleq 2 D\left(x_3\right),
 \end{align}

This implies that, if we choose $x_1 = 0$ and $x_2 = k_1^\star$, the previous inequality is satisfied for all points $x_3 \ge k_1^\star/2$, and in particular for $k_2^\star$.

This can be extended to a generic value of $r$ by repeatedly applying the convexity property of Eq.~\eqref{eq:convexity} for $r-1$ times and choosing $\lambda = 1/r, 1/(r-1), \dots$ until $\lambda\!=\!1/2$ in the last step. This process leads to:~
\begin{equation}
 (r-1) D(x_1) + D(x_2) \ge r D\left(\frac{r-1}{r}x_1 + \frac{1}{r} x_2\right).
\end{equation}
If we choose $x_1 = 0$ and $x_2 = k_1^\star$, and exploit Eq.~\eqref{eq:k_r}, we have that $(r-1)\,D(0) + D(k_1^\star) \ge r\,D(k_r^\star)$. 

\textbf{Step 3.} Finally, we show that the retransmission scheme leads to lower average distortion.

After $r$ consecutive transmissions, the data generated in the first slot has been either successfully sent or discarded, and the window of the packets involved in the processing-transmission mechanism moves forward. This means that focusing on $r$ transmissions suffices to encompass all possible situations.
There can be $2^r$ possible outcomes, each with a corresponding QoS cost.
The costs obtained in the two scenarios (retransmissions or single-transmission) differ only when a successful transmission follows $q < r$ consecutive failed transmissions: in this case, if no retransmission mechanism is considered, the total cost is given by $q D(0) + D(k_1^\star)$, otherwise it is $(q+1) D(k_{q+1}^\star)$ (see Section~\ref{sec:retx}).
We focused on the case $q=r-1$ and proved that $r\,D(k_r^\star) \ge (r\!-\!1)\,D(0) + D(k_1^\star)$. All other cases can be straightforwardly traced back to some $r' < r$ and the same result holds.
In practice, in all possible combinations of ACK/NACK in $r$ transmission attempts, the distortion obtained when the retransmission mechanism is adopted is not larger than that achieved in the single-transmission scenario. This completes the proof of the theorem.
\qed

\section{Proof of Theorem~\ref{th:retx_2}}\label{apx:proof_retx_2}

The distortion function $D(k)$ is convex and decreasing in $k\in\{0,\dots,m\}$.
We want to compare the performance obtained by the retransmission scheme with $r>2$ (case $A$) and with $r-1$ (case $B$) maximum number of retransmissions allowed.

As in the proof of Theorem~\ref{th:retx}, we neglect any energy constraint, so $k_r^\star = k_{r,R}^\star$ and $k_{r-1}^\star = k_{r-1,R}^\star$.
If we consider $r$ consecutive transmissions, the only difference between the two cases is obtained when a successful transmission follows $r-1$ failed attempts. 
The corresponding distortions are, respectively:
 \begin{align}
  D^A & = r D(k^\star_r) \\
  D^B & = (r-1)D(k^\star_{r-1}) + D(0).
 \end{align}
%
Since $0\le k^\star_{r} \le  k^\star_{r-1}$, it is $D(0) \ge  D(k^\star_{r}) \ge D(k^\star_{r-1})$.
We want to show that $D^A$ is not always lower than $D^B$, but under some circumstances it can be higher. 

If $k^\star_{r-1} > k^\star_r = 0$, then $D^A = r D(0) > (r-1) D(k^\star_{r-1}) + D(0) = D^B$.
Vice versa, if $k^\star_{r-1} = k^\star_r > 0$, then $D^A < D^B$.
This is more likely to happen when $r$ gets closer to $m$, which represents the maximum value that $k$ can take.
And, if $r>m$, it surely is $k^\star_q = k^\star_{q-1}$ for some $q\le r$.
\qed

\medskip

\begin{figure}
 \centering
 \setlength\fwidth{.85\columnwidth}
 \setlength\fheight{0.4\columnwidth}
 \definecolor{lavender}{rgb}{0.9020,0.9020,0.9804}%
\definecolor{lightskyblue}{rgb}{0.6784,0.8471,0.9020}%
\definecolor{deepskyblue}{rgb}{0,0.7490,1}%
\definecolor{steelblue}{rgb}{0.2745,0.5098,0.7059}%
\definecolor{blue}{rgb}{0,0,1}%
\definecolor{royalblue}{rgb}{0.2549,0.4118,0.8824}%

\definecolor{gainsboro}{rgb}{0.8627,0.8627,0.8627}%
\definecolor{darkslategrey}{rgb}{0.1843,0.3098,0.3098}%
\definecolor{gray}{rgb}{0.5,0.5,0.5}%
\definecolor{lightgray}{rgb}{0.77,0.77,0.77}%
\definecolor{midgray}{rgb}{0.6,0.6,0.6}%

\definecolor{orange}{rgb}{1,0.7,0.2}%
\definecolor{red}{rgb}{0.85,0,0.2}%

\definecolor{lightcoral}{rgb}{0.9412,0.5020,0.5020}%
\definecolor{indianred}{rgb}{0.8039,0.3608,0.3608}%
\definecolor{lightsalmon}{rgb}{1.0000,0.6275,0.4784}%
\definecolor{darksalmon}{rgb}{0.9137,0.5882,0.4784}%

\definecolor{mycolor2}{rgb}{0.85000,0.32500,0.09800}%
\definecolor{mycolor5}{rgb}{1.0000,0.6275,0.4784}%
\definecolor{mycolor3}{rgb}{0.92900,0.69400,0.12500}%
\definecolor{mycolor6}{rgb}{1.0000,0.9804,0.8039}%

\begin{tikzpicture}
\pgfplotsset{every tick label/.append style={font=\scriptsize}}
\tikzstyle{dotted}= [dash pattern=on \pgflinewidth off 0.5mm] 
\tikzstyle{dashed}= [dash pattern=on 7.5*0.57*0.8pt off 7.5*0.3*0.8pt] 

\begin{axis}[%
width=0.951\fwidth,
height=\fheight,
at={(0\fwidth,0\fheight)},
scale only axis,
xmin=0,
xmax=200,
xtick={0,21,140,200}, 
xticklabels={{ 0},{$k_V$},{$k^\star_{r-1}$},{$m$}}, 
xlabel style={font=\footnotesize\color{white!15!black}, at={(axis description cs:0.5,-0.07)},anchor=north},
xlabel={$k$},
ymin=0,
ymax=1,
ytick={0,0.23,1}, 
yticklabels={{ 0},{$D^B/r$},{1}}, 
ylabel style={font=\footnotesize\color{white!15!black}, at={(axis description cs:-0.1,0.5)},anchor=north},
ylabel={$D$},
axis background/.style={fill=white},
axis x line*=bottom,
axis y line*=left
]
\addplot [color=orange, line width=1.5pt, forget plot]
  table[row sep=crcr]{%
0	1\\
1	0.919345037967309\\
2	0.798362594918272\\
3	0.66640845862682\\
4	0.583514831809915\\
5	0.524725830611046\\
6	0.479985714823941\\
7	0.444323270145079\\
8	0.414948693406409\\
9	0.390153818169376\\
10	0.368823777934544\\
11	0.350194444124871\\
12	0.333721394549922\\
13	0.31900434378247\\
14	0.305741207562765\\
15	0.293698947140944\\
16	0.282694381771157\\
17	0.272581166527068\\
18	0.263240716950846\\
19	0.254575736629904\\
20	0.2465055065751\\
21	0.238962394687687\\
22	0.231889227513619\\
23	0.225237282586995\\
24	0.21896473477325\\
25	0.213035439688868\\
26	0.207417970773168\\
27	0.202084849591505\\
28	0.197011925006948\\
29	0.192177868235845\\
30	0.187563758977256\\
31	0.183152743754221\\
32	0.178929751983856\\
33	0.174881258552732\\
34	0.170995084124737\\
35	0.167260226268981\\
36	0.163666715920027\\
37	0.160205494782898\\
38	0.156868310151628\\
39	0.153647624281386\\
40	0.150536535984259\\
41	0.147528712539863\\
42	0.14461833034874\\
43	0.141800023027311\\
44	0.139068835862201\\
45	0.136420185719828\\
46	0.133849825652686\\
47	0.131353813563215\\
48	0.128928484384725\\
49	0.126570425320455\\
50	0.124276453749782\\
51	0.122043597467306\\
52	0.119869076968094\\
53	0.117750289532379\\
54	0.115684794896805\\
55	0.113670302327911\\
56	0.111704658937932\\
57	0.109785839103669\\
58	0.107911934866994\\
59	0.106081147210731\\
60	0.104291778116747\\
61	0.102542223324394\\
62	0.100830965717173\\
63	0.0991565692739782\\
64	0.0975176735285964\\
65	0.0959129884875714\\
66	0.0943412899620973\\
67	0.0928014152745123\\
68	0.0912922593042391\\
69	0.0898127708417831\\
70	0.0883619492227085\\
71	0.0869388412164374\\
72	0.0855425381472915\\
73	0.0841721732274847\\
74	0.0828269190837964\\
75	0.0815059854614543\\
76	0.0802086170903559\\
77	0.0789340917001843\\
78	0.0776817181722435\\
79	0.076450834816979\\
80	0.0752408077671653\\
81	0.0740510294776562\\
82	0.0728809173234112\\
83	0.0717299122882508\\
84	0.0705974777374537\\
85	0.0694830982679092\\
86	0.0683862786300745\\
87	0.0673065427164784\\
88	0.0662434326119496\\
89	0.0651965077011493\\
90	0.0641653438293522\\
91	0.0631495325127432\\
92	0.0621486801948036\\
93	0.0611624075456254\\
94	0.0601903488012465\\
95	0.0592321511403194\\
96	0.0582874740956381\\
97	0.0573559889982322\\
98	0.0564373784519104\\
99	0.0555313358362955\\
100	0.0546375648365331\\
101	0.0537557789979953\\
102	0.0528857013044136\\
103	0.0520270637779964\\
104	0.0511796071001813\\
105	0.0503430802517725\\
106	0.0495172401712965\\
107	0.0487018514304938\\
108	0.0478966859259336\\
109	0.0471015225858112\\
110	0.0463161470910488\\
111	0.0455403516098773\\
112	0.0447739345451353\\
113	0.044016700293566\\
114	0.0432684590164433\\
115	0.0425290264208997\\
116	0.041798223551367\\
117	0.0410758765905818\\
118	0.0403618166696374\\
119	0.0396558796865993\\
120	0.0389579061332298\\
121	0.0382677409293946\\
122	0.0375852332647504\\
123	0.0369102364473371\\
124	0.0362426077587201\\
125	0.0355822083153483\\
126	0.0349289029358142\\
127	0.0342825600137213\\
128	0.033643051395877\\
129	0.0330102522655524\\
130	0.0323840410305561\\
131	0.0317642992158927\\
132	0.0311509113607813\\
133	0.0305437649198271\\
134	0.0299427501681484\\
135	0.0293477601102729\\
136	0.0287586903926261\\
137	0.0281754392194453\\
138	0.0275979072719614\\
139	0.0270259976306987\\
140	0.0264596157007507\\
141	0.0258986691398981\\
142	0.0253430677894408\\
143	0.0247927236076243\\
144	0.0242475506055445\\
145	0.0237074647854227\\
146	0.0231723840811485\\
147	0.0226422283009901\\
148	0.022116919072381\\
149	0.0215963797886935\\
150	0.0210805355579142\\
151	0.0205693131531428\\
152	0.0200626409648362\\
153	0.0195604489547262\\
154	0.0190626686113419\\
155	0.0185692329070704\\
156	0.0180800762566927\\
157	0.0175951344773364\\
158	0.0171143447497864\\
159	0.0166376455811005\\
160	0.0161649767684776\\
161	0.0156962793643293\\
162	0.015231495642506\\
163	0.0147705690656359\\
164	0.0143134442535297\\
165	0.0138600669526126\\
166	0.0134103840063437\\
167	0.0129643433265841\\
168	0.0125218938658795\\
169	0.0120829855906207\\
170	0.0116475694550522\\
171	0.0112155973760934\\
172	0.0107870222089462\\
173	0.0103617977234577\\
174	0.00993987858121074\\
175	0.00952122031331662\\
176	0.00910577929888401\\
177	0.00869351274413971\\
178	0.00828437866217821\\
179	0.0078783358533176\\
180	0.00747534388604053\\
181	0.00707536307849924\\
182	0.00667835448056556\\
183	0.00628427985640647\\
184	0.00589310166756703\\
185	0.00550478305654385\\
186	0.00511928783083135\\
187	0.0047365804474257\\
188	0.00435662599777016\\
189	0.00397939019312745\\
190	0.00360483935036486\\
191	0.00323294037813788\\
192	0.00286366076345984\\
193	0.00249696855864412\\
194	0.00213283236860734\\
195	0.0017712213385214\\
196	0.00141210514180307\\
197	0.00105545396843029\\
198	0.000701238513574777\\
199	0.000349429966540774\\
200	0\\
};
\addplot [color=black, line width=0.2pt, forget plot]
  table[row sep=crcr]{%
140	0.0264596157007507\\
0	1\\
};
\addplot [color=black, line width=0.5pt, draw=none, mark=*, mark options={solid, fill=black, black}, forget plot]
  table[row sep=crcr]{%
110	0.237018903511996\\
};
\addplot [color=black, dashed, line width=1.0pt, forget plot]
  table[row sep=crcr]{%
21	0.237018903511996\\
110	0.237018903511996\\
};
\addplot [color=black, line width=0.5pt, draw=none, mark=*, mark options={solid, fill=black, black}, forget plot]
  table[row sep=crcr]{%
21	0.237018903511996\\
};
\addplot [color=black, dashed, line width=1.0pt, forget plot]
  table[row sep=crcr]{%
21	0\\
21	0.238962394687687\\
};
\addplot [color=red, line width=1.5pt, forget plot]
  table[row sep=crcr]{%
0	1\\
1	0.919345037967309\\
2	0.798362594918272\\
3	0.66640845862682\\
4	0.583514831809915\\
5	0.524725830611046\\
6	0.479985714823941\\
7	0.444323270145079\\
8	0.414948693406409\\
9	0.390153818169376\\
10	0.368823777934544\\
11	0.350194444124871\\
12	0.333721394549922\\
13	0.31900434378247\\
14	0.305741207562765\\
15	0.293698947140944\\
16	0.282694381771157\\
17	0.272581166527068\\
18	0.263240716950846\\
19	0.254575736629904\\
20	0.2465055065751\\
21	0.238962394687687\\
};
\addplot [color=black, line width=1pt, draw=none, mark=star, mark options={solid, fill=black, black}, forget plot]
  table[row sep=crcr]{%
140	0.0264596157007507\\
};
\addplot [color=black, line width=1pt, draw=none, mark=star, mark options={solid, fill=black, black}, forget plot]
  table[row sep=crcr]{%
0	1\\
};
\end{axis}
\end{tikzpicture}%
 \caption{Determination of $k_V$, see Eq.~\eqref{eq:k_V}. The red part of the curve represents distortion levels that are larger than $D_B/r$, the orange one is related to distortion levels that are smaller. If $k_r^\star$ falls to the right of $k_V$, then choosing $r-1$ rather than $r$ improves the performance.}
 \label{fig:proof_2}
\end{figure}
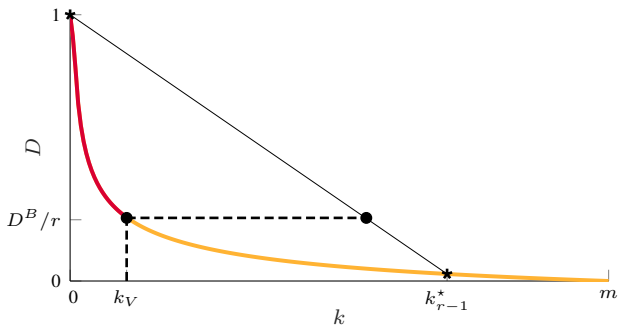

This suffices to prove the theorem, but in addition we want to understand what happens if $0 < k^\star_{r} < k^\star_{r-1}$.
We introduce the coefficient $\mu \triangleq (r-1)/r$, so that we compare $D_A/r = D(k^\star_r)$ and $D^B/r = \mu D(k^\star_{r-1}) + (1-\mu)D(0)$.
Let $k_V$ be the point such that $D(k_V) = D^B/r$, as represented in Fig.~\ref{fig:proof_2}.
By applying the definition of distortion given in Eq.~\eqref{eq:dist}, we have:~
\begin{equation} \label{eq:k_V}
 k_V = \left(\mu (k^\star_{r-1})^{-a} + (1-\mu) \frac{b+D_\fl}{b\,m^a} \right)^{-1/a}.
\end{equation}

The distortion is a decreasing function, hence $D(k)>D(k_V) = D^B/r$ for all $k < k_V$. This implies that:
\begin{equation}
 \begin{cases}
  D^A > D^B	&\quad \textrm{if } k^\star_r < k_V	\\
  D^A = D^B	&\quad \textrm{if } k^\star_r = k_V  	\\
  D^A < D^B	&\quad \textrm{if } k^\star_r > k_V
 \end{cases}
\end{equation}

Determining when $k^\star_r < k_V$ is a challenging task, hampered by the fact that there is no closed-form expression for the relationship between $r$ and $k^\star_r$, see Section~\ref{ssec:RDP} and~\cite{Pielli-2017}.
However, as $r$ increases, there is a higher probability that $D^A$ is larger than $D^B$. 
In particular, let $\mathcal{K}^- \triangleq \{k\in\mathbb{N}: 0<k<k_V\}$ and $\mathcal{K} \triangleq \{k\in\mathbb{N}: 0<k<k^\star_{r-1}\}$. 
As $r$ increases, $\mu = (r-1)/r \rightarrow 1$, and $k_V$ gets closer to $k^\star_{r-1}$ (from the left). Hence, the ratio $|\mathcal{K^-}|/|\mathcal{K}| \rightarrow 1$ and the probability that $k^\star_{r}$ falls in $\mathcal{K^-}$ rather than in $\mathcal{K}\setminus\mathcal{K^-}$ increases.

In practice, for small values of $r$, in general it holds that $k_V<k^\star_r < k^\star_{r-1}$, and increasing $r$ leads to lower distortion at the receiver.
But, as $r$ increases, $k_V$ gets very close to $k^\star_{r-1}$ and it is more unlikely that $k^\star_r$ falls in between these two values, thus increasing $r$ brings no benefit.

A curious behavior appears as $r$ gets closer to $m$, because the probability that $k_r=k_{r-1}$ becomes higher.
In this case, it may happen that (i) $k^\star_r < k_V$ and therefore it is better to select $r-1$ rather than $r$, but (ii) $k^\star_{r+1} = k^\star_r$ and therefore also $r+1$ is a better choice than $r$. 
Therefore, it may happen that both decreasing and increasing $r$ lead to a better performance (nevertheless, there is a single optimal choice).
Although in general the retransmission mechanism is effective, selecting the appropriate value of $r$ is not trivial, because the attainable gain is not proportional to $r$, but also depends on some system parameters such as the discretization of the compression ratio (hence, $m$) and the energy availability. 
This goes beyond the scope of the work.


\bibliographystyle{IEEEtran}
\bibliography{bibb}

\begin{thebibliography}{10}
\providecommand{\url}[1]{#1}
\csname url@samestyle\endcsname
\providecommand{\newblock}{\relax}
\providecommand{\bibinfo}[2]{#2}
\providecommand{\BIBentrySTDinterwordspacing}{\spaceskip=0pt\relax}
\providecommand{\BIBentryALTinterwordstretchfactor}{4}
\providecommand{\BIBentryALTinterwordspacing}{\spaceskip=\fontdimen2\font plus
\BIBentryALTinterwordstretchfactor\fontdimen3\font minus
  \fontdimen4\font\relax}
\providecommand{\BIBforeignlanguage}[2]{{%
\expandafter\ifx\csname l@#1\endcsname\relax
\typeout{** WARNING: IEEEtran.bst: No hyphenation pattern has been}%
\typeout{** loaded for the language `#1'. Using the pattern for}%
\typeout{** the default language instead.}%
\else
\language=\csname l@#1\endcsname
\fi
#2}}
\providecommand{\BIBdecl}{\relax}
\BIBdecl

\bibitem{Pielli-2017}
C.~Pielli, \v{C}. Stefanovi\'c, P.~Popovski, and M.~Zorzi, ``Minimizing data
  distortion of periodically reporting {IoT} devices with energy harvesting,''
  accepted for presentation at \emph{IEEE International Conference on Sensing,
  Communication and Networking (SECON), June} 2017.

\bibitem{Zorzi-2010}
M.~Zorzi, A.~Gluhak, S.~Lange, and A.~Bassi, ``From today's {INTRAnet} of
  things to a future {INTERnet} of things: a wireless- and mobility-related
  view,'' \emph{{IEEE Wireless} Communications}, vol.~17, no.~6, pp. 44--51,
  Dec. 2010.

\bibitem{nokia}
\BIBentryALTinterwordspacing
Nokia. (2015) {LTE evolution for IoT connectivity}. [Online]. Available:
  \url{http://resources.alcatel-lucent.com/asset/200178}
\BIBentrySTDinterwordspacing

\bibitem{ericsson}
\BIBentryALTinterwordspacing
Ericsson. (2015) {Ericsson, AT\&T and Altair demonstrate over 10 years of
  battery life on LTE IoT commercial chipset}. [Online]. Available:
  \url{https://www.ericsson.com/news/1962068}
\BIBentrySTDinterwordspacing

\bibitem{huawei}
\BIBentryALTinterwordspacing
Huawei. (2015) {NB-IOT - Enabling New Business Opportunities}. [Online].
  Available: \url{http://www.huawei.com/minisite/4-5g/img/NB-IOT.pdf}
\BIBentrySTDinterwordspacing

\bibitem{Uetal2015}
S.~Ulukus, A.~Yener, E.~Erkip, O.~Simeone, M.~Zorzi, P.~Grover, and K.~Huang,
  ``Energy harvesting wireless communications: A review of recent advances,''
  \emph{IEEE Journal on Selected Areas in Communications}, vol.~33, no.~3, pp.
  360--381, Mar. 2015.

\bibitem{Gunduz-2014}
D.~Gunduz, K.~Stamatiou, N.~Michelusi, and M.~Zorzi, ``Designing intelligent
  energy harvesting communication systems,'' \emph{{IEEE} Communications
  Magazine}, vol.~52, no.~1, pp. 210--216, Jan. 2014.

\bibitem{tutorial}
M.~L. Ku, W.~Li, Y.~Chen, and K.~J.~R. Liu, ``Advances in energy harvesting
  communications: Past, present, and future challenges,'' \emph{{IEEE
  Communications Surveys and Tutorials}}, vol.~18, no.~2, pp. 1384--1412,
  Second Quarter 2016.

\bibitem{schwartz-1993}
A.~Schwartz, ``A reinforcement learning method for maximizing undiscounted
  rewards,'' in \emph{Proceedings of the tenth international conference on
  machine learning}, vol. 298, 1993, pp. 298--305.

\bibitem{Sutton-98}
R.~S. Sutton and A.~G. Barto, \emph{Reinforcement learning: An
  introduction}.\hskip 1em plus 0.5em minus 0.4em\relax MIT press Cambridge,
  1998, vol.~1, no.~1.

\bibitem{Blasco-2013}
P.~Blasco, D.~Gunduz, and M.~Dohler, ``A learning theoretic approach to energy
  harvesting communication system optimization,'' \emph{IEEE Transactions on
  Wireless Communications}, vol.~12, no.~4, pp. 1872--1882, Apr. 2013.

\bibitem{Ortiz-2016}
A.~Ortiz, H.~Al-Shatri, X.~Li, T.~Weber, and A.~Klein, ``Reinforcement learning
  for energy harvesting point-to-point communications,'' in \emph{{IEEE}
  International Conference on Communications {(ICC)}}, May 2016, pp. 1--6.

\bibitem{Zachariadis}
K.~E. Zachariadis, M.~L. Honig, and A.~K. Katsaggelos, ``Source fidelity over
  fading channels: performance of erasure and scalable codes,'' \emph{{IEEE
  Transactions on Communications}}, vol.~56, no.~7, pp. 1080--1091, July 2008.

\bibitem{Aguerri-2011}
I.~E. Aguerri and D.~Gunduz, ``Expected distortion with fading channel and side
  information quality,'' in \emph{{IEEE} International Conference on
  Communications (ICC)}, June 2011, pp. 1--6.

\bibitem{Laneman}
J.~N. Laneman, E.~Martinian, G.~W. Wornell, and J.~G. Apostolopoulos,
  ``Source-channel diversity for parallel channels,'' \emph{{IEEE Transactions
  on Information Theory}}, vol.~51, no.~10, pp. 3518--3539, Oct. 2005.

\bibitem{Zhao-2016}
S.~Zhao, D.~Tuninetti, R.~Ansari, and D.~Schonfeld, ``On achievable distortion
  exponents for a gaussian source transmitted over parallel gaussian channels
  with correlated fading and asymmetric {SNR}s,'' \emph{IEEE Transactions on
  Information Theory}, vol.~62, no.~7, pp. 4135--4153, July 2016.

\bibitem{etemadi}
F.~Etemadi and H.~Jafarkhani, ``A unified framework for layered transmission
  over fading and packet erasure channels,'' \emph{{IEEE Transactions on
  Communications}}, vol.~56, no.~4, pp. 565--573, Apr. 2008.

\bibitem{ng-2014}
D.~W.~K. Ng, R.~Schober, and H.~Alnuweiri, ``Secure layered transmission in
  multicast systems with wireless information and power transfer,'' in
  \emph{2014 {IEEE} International Conference on Communications {(ICC)}}, June
  2014, pp. 5389--5395.

\bibitem{bhat}
R.~V. Bhat, M.~Motani, and T.~J. Lim, ``Distortion minimization in energy
  harvesting sensor nodes with compression power constraints,'' in \emph{{2016
  IEEE International Conference on Communications (ICC)}}, May 2016, pp. 1--6.

\bibitem{Castiglione}
P.~Castiglione, O.~Simeone, E.~Erkip, and T.~Zemen, ``Energy management
  policies for energy-neutral source-channel coding,'' \emph{{IEEE Transactions
  on Communications}}, vol.~60, no.~9, pp. 2668--2678, Sept. 2012.

\bibitem{motlagh}
M.~S. Motlagh, M.~B. Khuzani, and P.~Mitran, ``On lossy joint source-channel
  coding in energy harvesting communication systems,'' \emph{{IEEE Transactions
  on Communications}}, vol.~63, no.~11, pp. 4433--4447, Nov. 2015.

\bibitem{bui}
N.~Bui and M.~Rossi, ``Staying alive: System design for self-sufficient sensor
  networks,'' \emph{ACM Transactions on Sensor Networks}, vol.~11, no.~3, pp.
  40:1--40:42, Feb. 2015.

\bibitem{zordan}
D.~Zordan, T.~Melodia, and M.~Rossi, ``On the design of temporal compression
  strategies for energy harvesting sensor networks,'' \emph{{IEEE Transactions
  on Wireless Communications}}, vol.~15, no.~2, pp. 1336--1352, Feb. 2016.

\bibitem{Fullana-2017}
M.~Calvo-Fullana, J.~Matamoros, and C.~Antón-Haro, ``Reconstruction of
  correlated sources with energy harvesting constraints in delay-constrained
  and delay-tolerant communication scenarios,'' \emph{{IEEE} Transactions on
  Wireless Communications}, vol.~16, no.~3, pp. 1974--1986, Mar. 2017.

\bibitem{aprem-2013}
A.~Aprem, C.~R. Murthy, and N.~B. Mehta, ``Transmit power control policies for
  energy harvesting sensors with retransmissions,'' \emph{{IEEE} Journal of
  Selected Topics in Signal Processing}, vol.~7, no.~5, pp. 895--906, Oct.
  2013.

\bibitem{Costa-2013}
D.~G. Costa, L.~A. Guedes, F.~Vasques, and P.~Portugal, ``Partial
  energy-efficient hop-by-hop retransmission in wireless sensor networks,'' in
  \emph{11th {IEEE} International Conference on Industrial Informatics
  {(INDIN)}}, July 2013, pp. 146--151.

\bibitem{Hu-2013}
N.~Hu, Y.~D. Yao, and Z.~Yang, ``Analysis of cooperative {TDMA} in rayleigh
  fading channels,'' \emph{{IEEE} Transactions on Vehicular Technology},
  vol.~62, no.~3, pp. 1158--1168, Mar. 2013.

\bibitem{Lee-2012}
J.~K. Lee, H.~J. Noh, and J.~Lim, ``Dynamic cooperative retransmission scheme
  for {TDMA} systems,'' \emph{{IEEE} Communications Letters}, vol.~16, no.~12,
  pp. 2000--2003, Dec. 2012.

\bibitem{lorawan_specs}
N.~Sornin, M.~Luis, T.~Eirich, T.~Kramp, and O.~Hersent, ``{LoRaWAN
  Specifications},'' LoRa Alliance, Tech. Rep., 2015.

\bibitem{Zordan-2017}
D.~Zordan, M.~Rossi, and M.~Zorzi, ``Rate-distortion classification for
  self-tuning {IoT} networks,'' \emph{{IEEE} International Conference on
  Communications {(ICC)}}, May 2017.

\bibitem{eccentric}
A.~Biason, C.~Pielli, M.~Rossi, A.~Zanella, D.~Zordan, M.~Kelly, and M.~Zorzi,
  ``{EC-CENTRIC}: An {Energy-} and {Context-Centric} perspective on {IoT}
  systems and protocol design,'' \emph{IEEE Access}, 2017.

\bibitem{polyansky}
Y.~Polyanskiy, H.~V. Poor, and S.~Verdu, ``Channel coding rate in the finite
  blocklength regime,'' \emph{IEEE Transactions on Information Theory},
  vol.~56, no.~5, pp. 2307--2359, May 2010.

\bibitem{yang}
W.~Yang, G.~Durisi, T.~Koch, and Y.~Polyanskiy, ``Quasi-static multiple-antenna
  fading channels at finite blocklength,'' \emph{IEEE Transactions on
  Information Theory}, vol.~60, no.~7, pp. 4232--4265, July 2014.

\bibitem{Rappaport-1996}
T.~S. Rappaport \emph{et~al.}, \emph{Wireless communications: principles and
  practice}.\hskip 1em plus 0.5em minus 0.4em\relax Prentice Hall {PTR} New
  Jersey, 1996, vol.~2.

\bibitem{Zordan-2014}
D.~Zordan, B.~Martinez, I.~Vilajosana, and M.~Rossi, ``On the performance of
  lossy compression schemes for energy constrained sensor networking,''
  \emph{ACM Transactions on Sensor Networks}, vol.~11, no.~1, pp. 15:1--15:34,
  Nov. 2014.

\bibitem{Howard-2006}
S.~L. Howard, C.~Schlegel, and K.~Iniewski, ``Error control coding in low-power
  wireless sensor networks: When is {ECC} energy-efficient?'' \emph{EURASIP
  Journal on Wireless Communication Networks}, vol. 2006, no.~2, pp. 1--14,
  Apr. 2006.

\bibitem{Ku}
M.~L. Ku, Y.~Chen, and K.~J.~R. Liu, ``Data-driven stochastic models and
  policies for energy harvesting sensor communications,'' \emph{{IEEE Journal
  on Selected Areas in Communications}}, vol.~33, no.~8, pp. 1505--1520, Aug.
  2015.

\bibitem{solarstat}
M.~Miozzo, D.~Zordan, P.~Dini, and M.~Rossi, ``Solarstat: Modeling photovoltaic
  sources through stochastic {Markov} processes,'' in \emph{2014 IEEE
  International Energy Conference}, May 2014, pp. 688--695.

\bibitem{altman}
E.~Altman, \emph{Constrained Markov decision processes}.\hskip 1em plus 0.5em
  minus 0.4em\relax CRC Press, 1999, vol.~7.

\bibitem{bertsekas}
D.~Bertsekas, \emph{{Dynamic Programming and Optimal Control}}, 4th~ed.\hskip
  1em plus 0.5em minus 0.4em\relax Belmont, MA,USA: Athena Scientific, 2012,
  vol.~2.

\end{thebibliography}

\end{document}